\documentclass[superscriptaddress,reprint,onecolumn,showkeys]{revtex4-2}
\usepackage{listings}
\usepackage{graphicx}% Include figure files
\usepackage{dcolumn}% Align table columns on decimal point
\usepackage{bm}% bold math
\usepackage{braket}

\usepackage[dvipsnames]{xcolor}
\usepackage{tikz}
\usepackage{amsmath}

\usepackage{makecell}
\usepackage{qcircuit}
\usepackage{subfig}
\usepackage{float}
\usepackage{booktabs}

\usepackage[justification=raggedright]{caption}
 
\usepackage[colorlinks,linkcolor=blue,anchorcolor=blue,citecolor=blue,urlcolor=blue]{hyperref}

\begin{document}

\title{A Modular Benchmark of Variational Quantum Attack Algorithms for S-DES}

\author{Zeguo Wang}
\affiliation{Quantum Science Center of Guangdong-Hong Kong-Macao Greater Bay Area, Shenzhen, China}

 \author{Quanfeng Lu}
 \affiliation{State Key Laboratory of Low-Dimensional Quantum Physics and Department of Physics, Tsinghua University, Beijing 100084, China}
 \author{Wentao Yang}
 \affiliation{State Key Laboratory of Low-Dimensional Quantum Physics and Department of Physics, Tsinghua University, Beijing 100084, China}
\author{Shijie Wei}
\email{weisj@baqis.ac.cn}
\affiliation{Beijing Academy of Quantum Information Sciences, Beijing 100193, China}
\author{Gui-Lu Long}
\email{gllong@tsinghua.edu.cn}
\affiliation{State Key Laboratory of Low-Dimensional Quantum Physics and Department of Physics, Tsinghua University, Beijing 100084, China \\
Beijing Academy of Quantum Information Sciences, Beijing 100193, China}

\author{Kai Wen}
\email{wenkai@quantumsc.cn}
\affiliation{Quantum Science Center of Guangdong-Hong Kong-Macao Greater Bay Area, Shenzhen, China}
\thanks{Corresponding Author: weisj@baqis.ac.cn, gllong@tsinghua.edu.cn, wenkai@quantumsc.cn}

\date{\today}
\begin{abstract}
Variational quantum algorithms (VQAs) have emerged as a promising approach to quantum cryptanalysis on noisy intermediate-scale quantum (NISQ) devices. Although numerous variational attack schemes have been proposed for symmetric cryptosystems, a systematic and modular benchmarking framework to evaluate their performance is still lacking.
In this work, we present a comprehensive benchmark study of variational quantum attacks on the Simplified Data Encryption Standard (S-DES), focusing on the modular design choices that determine attack efficiency. We formulate variational quantum attacks within a unified framework consisting of four components: initial state preparation, parameterized circuit (Ansatz) design, cost function construction, and classical optimization. Through numerical simulations, we systematically compare representative design alternatives and evaluate their combinations in terms of convergence behavior, success probability, and effective time complexity. We further introduce standardized metrics for assessing variational quantum attack performance.
Our results reveal clear performance hierarchies among different modular configurations and show that carefully optimized designs can significantly outperform naive quantum search. This work establishes a principled benchmark methodology for variational quantum cryptanalysis and positions S-DES as a practical testbed for evaluating quantum attacks on symmetric ciphers in the NISQ era.
\end{abstract}

\keywords{Quantum computation, Quantum algorithm, Cryptography}

\maketitle

\section{Introduction}\label{Sec1}

Quantum algorithms pose a long-term threat to modern cryptography, with Shor's algorithm~\cite{shor1994algorithms,gidney2025factor} endangering public-key systems~\cite{rivest1978method} and Grover's algorithm~\cite{grover1996fast,long2001grover} providing a quadratic speedup for symmetric-key search. While the implications of Grover's algorithm for block ciphers are well understood in theory, its practical realization requires fault-tolerant quantum hardware with substantial qubit and gate resources~\cite{wang2022quantum,li2023new}. This gap between theoretical asymptotic speedup and near-term implementability has motivated the exploration of alternative quantum attack paradigms suitable for noisy intermediate-scale quantum (NISQ)~\cite{preskill2018quantum} devices.

Variational quantum algorithms (VQAs)~\cite{peruzzo2014variational,cerezo2021variational,farhi2014quantum}, which combine parameterized quantum circuits with classical optimization loops, have emerged as a leading candidate for near-term quantum advantage. Originally developed for quantum chemistry~\cite{mcardle2020quantum,wen2024full} and many-body physics~\cite{fauseweh2024quantum}, VQAs have recently been adapted to cryptanalytic tasks, including key recovery attacks on symmetric ciphers~\cite{wang2022variational,wang2025reducing,aizpurua2025hacking}. In this variational quantum attack framework, the key search problem is encoded into a cost function whose minimum corresponds to the correct secret key, and the optimization process iteratively increases the overlap between the variational quantum state and the target key state.

Despite rapid progress, existing variational quantum attacks are often presented as problem-specific constructions, making it difficult to assess their relative performance or to identify which design choices are essential for achieving practical speedups. In particular, the performance of a variational quantum attack is highly sensitive to several modular components, including the choice of initial state, the structure and depth of the parameterized circuit (Ansatz), the form of the cost function, and the classical optimization algorithm. A lack of standardized benchmarks has hindered a clear understanding of how these components interact and how variational attacks compare with conventional quantum search strategies such as Grover's algorithm.

In this work, we address this gap by introducing a modular benchmark framework for variational quantum attack algorithms, using the Simplified Data Encryption Standard (S-DES)\cite{schaefer1996simplified} as a representative testbed. S-DES preserves the essential structural features of real-world block ciphers while remaining amenable to exhaustive classical simulation, making it well suited for systematic algorithmic benchmarking. Rather than proposing a single optimized attack circuit, we decompose variational quantum attacks into well-defined modules and perform a comparative evaluation across a wide range of design choices.

Our contributions are threefold. First, we introduce a unified modular formulation of variational quantum attacks on symmetric ciphers, explicitly characterizing the functional role and interaction of each algorithmic component.
Second, we perform extensive large-scale classical simulations to systematically benchmark different module combinations with respect to convergence behavior, success probability, and effective time complexity.
Third, we distill practical design principles that identify the conditions under which variational quantum attacks can surpass naive quantum search strategies under realistic resource constraints.
More broadly, this work establishes a principled and reproducible methodology for evaluating and comparing variational quantum cryptanalysis schemes. It lays the foundation for future experimental implementations on NISQ hardware and provides a scalable framework for extending benchmark-driven analysis to larger and more realistic cryptographic primitives.

% Our contributions are threefold. First, we provide a unified modular formulation of variational quantum attacks on symmetric ciphers, clarifying the role of each algorithmic component. Second, we conduct extensive classical simulations to benchmark different module combinations in terms of convergence speed, success probability, and effective time complexity. Third, we extract practical design guidelines that highlight when and how variational quantum attacks can outperform naive quantum search under realistic resource constraints.

% This work aims to establish a principled methodology for evaluating and comparing variational quantum cryptanalysis schemes, and to lay the groundwork for future implementations on NISQ hardware as well as extensions to larger and more realistic cryptographic primitives.

\section{Variational Quantum Attack Algorithm}\label{sec_II}

In this section, we introduce a unified framework for variational quantum attacks on symmetric cryptosystems. Our goal is not to present a single optimized attack circuit, but rather to formalize the problem setting and provide a modular abstraction that enables systematic benchmarking of different variational design choices. The framework described here serves as the foundation for all subsequent analysis and numerical experiments.

\subsection{Problem Formulation and Threat Model}

We consider a symmetric-key block cipher with a fixed plaintext--ciphertext pair $(P, C)$ and an unknown secret key $k^\ast$ drawn from a finite key space $\mathcal{K} = \{0,1\}^n$. The encryption function is denoted by
$C = E_{k^\ast}(P)$, 
where $E_k(\cdot)$ represents the encryption algorithm parameterized by the secret key $k$. 
It is worth noting that, for a given set of plaintext--ciphertext pairs, the key $k^\ast$ is not necessarily unique. In such cases, Ref.~\citep{KJPL202504001} employs multiple plaintext--ciphertext pairs simultaneously to uniquely determine the correct key. 
In contrast, we assume that only a single plaintext--ciphertext pair is available, which generally leads to multiple candidate keys consistent with the observed data. We define the success probability as the sum of the probabilities associated with all such candidate keys appearing in the experimental outcomes.
When additional ciphertexts are subsequently intercepted, these candidate keys can be tested sequentially by attempting plaintext reconstruction. The correct key can then be identified based on whether the decrypted plaintext yields meaningful information. In this way, our attack scheme requires only one plaintext--ciphertext pair.

Two variational quantum attack algorithms (VQAAs) for symmetric cryptosystems are proposed in Ref.~\citep{wang2022variational} and Refs.~\citep{wang2025reducing,aizpurua2025hacking}, respectively. In Ref.~\citep{wang2022variational}, the encryption process is implemented on a quantum computer, and the resulting superposition of ciphertexts is measured and transferred to a classical computer for cost function evaluation. In contrast, Refs.~\citep{wang2025reducing,aizpurua2025hacking} perform the encryption on a classical computer by combining the known plaintext with the superposition of keys prepared on the quantum device, and subsequently evaluate the cost function classically. The equivalence of these two VQAA models has been rigorously proven. A brief explanation is provided below.

The quantum encryption circuit, composed of Toffoli, CNOT, and X gates, implements a reversible classical transformation and maps each computational basis state to another without generating superpositions between distinct basis states. The quantum state prepared by the ansatz can be written as
\begin{equation}
\ket{\psi(\theta)} = \sum_i c_i \ket{x_i},
\end{equation}
and let the encryption function be $y_i = f(x_i)$. After applying the quantum encryption circuit, the state becomes
\begin{equation}
\ket{\psi'(\theta)} = \sum_i c_i \ket{y_i}.
\end{equation}
If the mapping $f$ is one-to-one, the probability of observing $y_i$ after encryption is $|c_i|^2$, which is identical to the probability of observing $x_i$ before encryption. Therefore, the same output distribution can be reproduced by applying the classical encryption function $f$ to samples drawn from the distribution $\{|c_i|^2\}$.
More generally, if multiple inputs $x_{i(1)}, x_{i(2)}, \ldots, x_{i(s)}$ satisfy $f(x_{i(k)}) = y_i$, then the probability of obtaining $y_i$ is given by
\begin{equation}
|c_{i(1)}|^2 + |c_{i(2)}|^2 + \cdots + |c_{i(s)}|^2,
\end{equation}
regardless of whether the encryption is performed on a quantum or classical computer. This establishes the equivalence between the two VQAAs.

In this work, we focus on the Simplified Data Encryption Standard (S-DES) as a representative benchmark cipher, although the framework is not limited to this specific construction. The details of S-DES can be found in Appendix~\ref{appen_A}.
A variational quantum attack is formulated as a hybrid quantum--classical optimization process. The secret key register is encoded into a quantum state, and a parameterized quantum circuit prepares a variational state
$\ket{\psi(\boldsymbol{\theta})} = U(\boldsymbol{\theta}) \ket{\psi_0},$ where $\ket{\psi_0}$ is an initial state and $U(\boldsymbol{\theta})$ denotes a parameterized quantum circuit with tunable parameters $\boldsymbol{\theta}$. Measurement outcomes obtained from $\ket{\psi(\boldsymbol{\theta})}$ are evaluated through a cost function designed such that its minimum corresponds to the correct key $k^\ast$. A classical optimizer then updates $\boldsymbol{\theta}$ iteratively based on the measured cost. We assume a noisy intermediate-scale quantum (NISQ) regime, where circuit depth and coherence time are limited, and fault-tolerant error correction\citep{kecceci2025quantum} is not available.

Under this threat model, the performance of a quantum attack is evaluated not only by asymptotic query complexity, but also by practical metrics such as convergence speed, success probability, circuit depth, and measurement overhead. These considerations motivate the use of variational quantum algorithms as a flexible attack paradigm that can trade off quantum resources against classical post-processing.

\subsection{Modular Decomposition of Variational Quantum Attacks}

Within the above problem setting, a variational quantum attack can be abstracted as an iterative hybrid optimization process. Starting from an initial quantum state, a parameterized quantum circuit prepares a trial state whose measurement statistics are evaluated through a cost function derived from the cryptographic objective. Classical optimization then updates the circuit parameters to progressively increase the overlap between the variational state and the target key state.
% \begin{enumerate}
%     \item \textbf{Initial state preparation,}
%     \item \textbf{Parameterized circuit (Ansatz) design,}
%     \item \textbf{Cost function construction,}
%     \item \textbf{Classical optimization strategy.}
% \end{enumerate}

\begin{figure}[h]
	\includegraphics[width=0.8\linewidth]{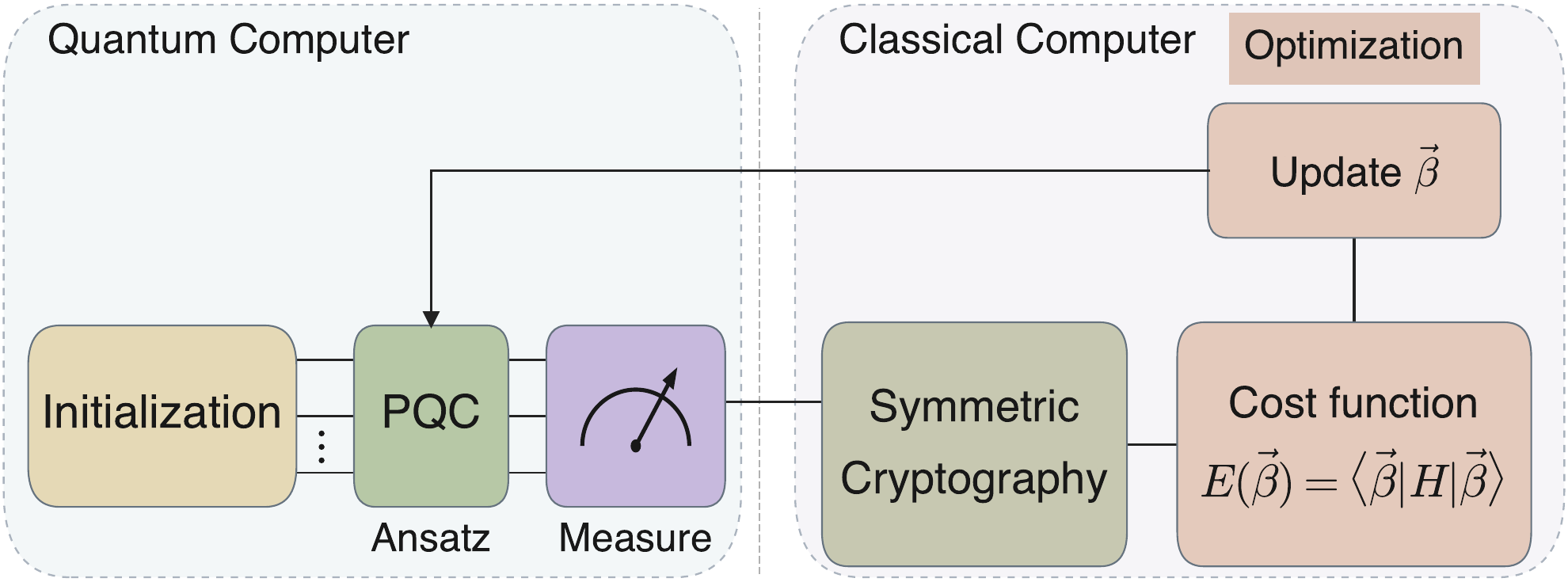}
	\caption{\textbf{Modular framework of variational quantum attacks.}
A variational quantum attack is decomposed into four core modules: (i) initial state preparation, (ii) parameterized quantum circuit (Ansatz) design, (iii) cost function construction, and (iv) classical optimization strategy. The quantum processor prepares a parameterized state and returns measurement outcomes used to evaluate the cost function, while a classical optimizer iteratively updates circuit parameters. This modular abstraction enables systematic benchmarking of different design choices and provides a unified basis for comparing variational attack configurations.
}\label{fig:framework}
\end{figure}

We decompose a variational quantum attack into four core modules: initial state preparation, parameterized circuit (Ansatz) design, cost function construction, and classical optimization strategy.
As shown in Fig.~\ref{fig:framework}, each module can be designed and optimized independently, while their interplay ultimately determines the overall attack performance.

The initial state preparation  defines the starting point of the optimization landscape. Typical choices include the uniform superposition over all candidate keys, which guarantees nonzero overlap with the correct key state, as well as Grover-enhanced states that amplify promising subspaces before variational optimization.
The parameterized quantum circuit (PQC), or Ansatz, governs the expressibility of the variational state and the reachable region of the Hilbert space. Shallow, hardware-efficient Ansatze are favored for NISQ implementations, while deeper or non-unitary-inspired constructions can improve expressibility at the cost of additional resources. The Ansatz structure directly impacts both trainability and robustness to noise.
The cost function encodes the cryptanalytic objective into a measurable quantity. In the context of symmetric key recovery, the cost is typically designed such that its global minimum corresponds to the correct secret key. Examples include graph-based Hamiltonians derived from logical constraints and Hamming-distance-based costs that compare trial ciphertexts with the known target ciphertext.
Finally, the classical optimization strategy determines how circuit parameters are updated based on measurement outcomes. Gradient-based methods can offer fast convergence on smooth landscapes, whereas gradient-free methods are often more robust to noise and measurement fluctuations. The choice of optimizer plays a crucial role in balancing convergence speed and stability.

By isolating these four modules, the proposed framework enables a systematic exploration of the design space of variational quantum attacks. In the following sections, we benchmark representative choices for each module using classical simulations and analyze how different combinations affect the efficiency and reliability of the attack.

\section{Design of Core Modules}\label{sec_III}

Within the variational quantum attack framework described above, overall performance is governed by a set of interdependent design choices. To enable systematic benchmarking, we decompose variational quantum attacks into several core modules and analyze their design principles separately. In this section, we focus on four key modules that directly shape the optimization landscape and the efficiency of key recovery: initialization strategies, Ansatz design, cost function construction and classical optimization strategy.

\subsection{Initialization Strategies}

The initialization strategy determines the starting point of the variational optimization and plays a crucial role in both convergence speed and robustness. In symmetric key search problems, the target key state typically lacks exploitable algebraic or physical structure, making informed initialization particularly challenging. A natural and widely adopted choice is the uniform superposition state over the entire key space
\begin{equation}
\ket{\psi_0} = \frac{1}{\sqrt{|\mathcal{K}|}} \sum_{k \in \mathcal{K}} \ket{k},
\end{equation}
which guarantees nonzero overlap with the correct key state. This strategy is unbiased and simple to implement, but it may lead to slow convergence due to the exponentially large search space.

To mitigate this issue, Grover-enhanced initialization strategies can be employed. By applying a small number of Grover iterations prior to variational optimization, the amplitude associated with candidate keys consistent with the plaintext-ciphertext relation can be partially amplified. The Grover iteration can be expressed as $\ket{\psi_1} = G^n \ket{\psi_0}$, 
where \(1 \leq n \leq \frac{\pi}{4}\sqrt{N/M}\). Here, \(N\) denotes the dimension of the search space, and \(M\) is the number of candidate keys.
The Grover operator is defined as
\begin{equation}
G = \left( 2\ket{\psi_0}\bra{\psi_0} - I \right) O ,
\end{equation}
where \(O\) is the oracle implemented via the quantum realization of the S-DES encryption circuit together with its inverse, combined with a multi-qubit controlled operation. The detailed construction of the oracle can be found in Ref.~\cite{wang2022variational}.
Such hybrid initialization schemes preserve the generality of the variational framework while improving the effective overlap with the target state. We expect them to facilitate faster convergence in subsequent optimization steps.

\subsection{Ansatz Design}

The parameterized quantum circuit, or Ansatz, defines the family of quantum states accessible during the optimization process. Its structure directly affects expressibility, trainability, and resource requirements. In the context of variational quantum attacks, the Ansatz must balance the ability to represent the target key state against the limitations imposed by NISQ hardware.

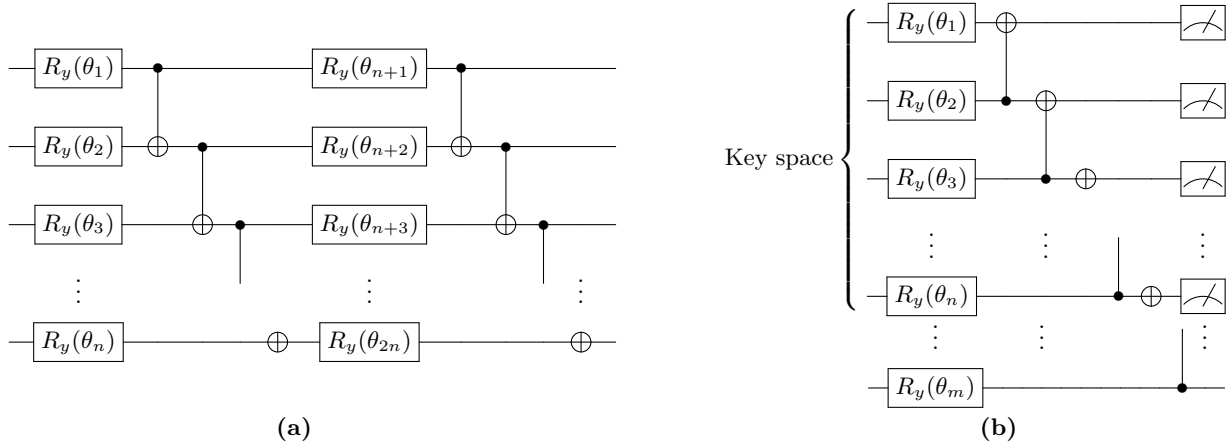
\begin{figure}[h]
\centering

% ---------- Circuits ----------
\begin{minipage}[c]{0.48\textwidth}
\centering
\[
\Qcircuit @C=1.0em @R=1.6em {
\lstick{} & \gate{R_y(\theta_1)}   & \ctrl{1} & \qw   & \qw   & \qw      & \gate{R_y(\theta_{n+1})} & \ctrl{1} & \qw   & \qw   & \qw  & \qw \\
\lstick{} & \gate{R_y(\theta_2)}   & \targ    & \ctrl{1} & \qw     & \qw & \gate{R_y(\theta_{n+2})} & \targ    & \ctrl{1} & \qw & \qw & \qw \\
\lstick{} & \gate{R_y(\theta_3)}   & \qw      & \targ    & \ctrl{1} & \qw & \gate{R_y(\theta_{n+3})} & \qw      & \targ    & \ctrl{1} & \qw & \qw \\
\lstick{} & \vdots         &          & & &       & \vdots &                          &        &       & \vdots \\
\lstick{} & \gate{R_y(\theta_n)}   & \qw      & \qw   & \qw   & \targ    & \gate{R_y(\theta_{2n})}  & \qw      & \qw   & \qw   & \targ & \qw
}
\]
\end{minipage}
\hspace{0.001\textwidth}
\begin{minipage}[c]{0.48\textwidth}
\centering
\[
\begin{array}{l}

\text{Key space}
\left\{
\begin{array}{l}
\Qcircuit @C=0.8em @R=1.6em {
\lstick{} & \gate{R_y(\theta_1)} & \targ    & \qw      & \qw      & \qw   & \qw   & \meter \\
\lstick{} & \gate{R_y(\theta_2)} & \ctrl{-1}& \targ    & \qw      & \qw   & \qw   & \meter \\
\lstick{} & \gate{R_y(\theta_3)} & \qw      & \ctrl{-1}& \targ    & \qw   & \qw   & \meter \\
\lstick{} & \vdots          &               & \vdots &                                     &  & & \vdots \\
\lstick{} & \gate{R_y(\theta_n)} & \qw      & \qw   & \qw   & \ctrl{-1}& \targ    & \meter \\
}
\end{array}
\right.

\\[1em]

\hspace{5.85em}  

\Qcircuit @C=0.8em @R=1.6em {
\lstick{} &  \vdots                          &&&       \vdots   & & & & &&&& \vdots      \\
\lstick{} & \gate{R_y(\theta_m)} & \qw      & \qw   & \qw & \qw  & \qw    & \qw  & \qw & \qw & \qw    & \ctrl{-1} & \qw & \qw 
}

\end{array}
\]
\end{minipage}

\vspace{0.3em}

% ---------- Labels ----------
\begin{minipage}{0.48\textwidth}
\centering
\textbf{(a)}
\end{minipage}
\hfill
\begin{minipage}{0.48\textwidth}
\centering
\textbf{(b)}
\end{minipage}

\caption{Two Ansatz architectures.
(a) A fully unitary Ansatz composed of alternating single-qubit $R_y(\theta_i)$ rotations and C-NOT gates;The circuit depth is controlled by the number of repeated layers and determines the expressibility of the variational state.
(b) A non-unitary-inspired Ansatz incorporating ancillary qubits and mid-circuit measurements. The first $n$ qubits encode candidate key states and the remaining $m-n$ qubits  induce effective non-unitary dynamics through measurement-induced state reduction. This architecture enhances expressive power by emulating dissipative or contractive transformations at the cost of additional quantum resources.
}
\label{fig:ansatz}

\end{figure}

A common approach is to employ layered unitary Ansatze composed of alternating single-qubit rotations and entangling gates which is shown in Fig~\ref{fig:ansatz}(a).  Such constructions are flexible and hardware-compatible, with circuit depth serving as a tunable parameter controlling expressibility. However, overly deep circuits may suffer from noise accumulation and barren plateau phenomena~\cite{larocca2025barren,qi2023barren}, degrading optimization performance.

Beyond strictly unitary designs, non-unitary-inspired Ansatze showing in Fig.~\ref{fig:ansatz}(b) offer an alternative route to enhancing expressibility. By introducing ancillary qubits and effective non-unitary transformations through partial tracing or post-selection, these Ansatze can emulate dissipative or contractive dynamics~\cite{weimer2021simulation}. While they incur additional overhead, non-unitary-inspired designs have the potential to reshape the optimization landscape and reduce the difficulty of locating the global minimum associated with the correct key.

\subsection{Cost Function Construction} \label{sec:cost_function}

The cost function provides the interface between the quantum state preparation and the classical optimization loop. Its role is to encode the cryptanalytic objective into a measurable quantity whose minimum corresponds to successful key recovery. An effective cost function should exhibit a smooth optimization landscape, be resilient to measurement noise, and remain efficiently evaluable on NISQ devices. 

One class of cost functions is derived from graph-based or Hamiltonian formulations, where logical constraints imposed by the encryption algorithm are mapped onto an energy function. In this formulation, the \textcolor{red}{known ciphertext} corresponds to the ground state of the effective Hamiltonian, and minimizing the expected energy guides the variational state toward the solution. The Hamiltonian can be expressed as follows:
\begin{equation}
H=\sum_{\{i,j\} \in E} w_{ij} Z_i Z_j + \sum_{i=0}^7 t_i Z_i
\end{equation}
where $E$ is the set of connected node pairs in the graph, $i$ and $j$ are node indices, $w_{ij}$ denotes the weight of the edge connecting nodes $i$ and $j$, and $Z$ is the Pauli-$Z$ operator.

The coefficients $w_{i,j}$ and $t_i$ are defined as:
\begin{equation}
w_{ij} =
\begin{cases}
a, & \text{if } V(i) \neq V(j), \\
-a, & \text{if } V(i) = V(j).
\end{cases}
\end{equation}

\begin{equation}
t_i =
\begin{cases}
b, & \text{if } V(i) = 1, \\
-b, & \text{if } V(i) = 0.
\end{cases}
\end{equation}
where $V(i)$ represents the value of the $i$-th bit of the ciphertext. The quantity $r$ is defined as the ratio between the energy gap separating the ground state from the first excited state and the total spectral width. Empirically, larger values of $r$ are observed to correlate with improved convergence behavior in our benchmark setting, as demonstrated by the experimental results presented in Appendix~\ref{appen_B}. Proportional scaling of $a$ and $b$ does not alter the relative structure of the optimization landscape, as it uniformly rescales the Hamiltonian spectrum. Accordingly, $b = 1$ is fixed throughout, and only $a$ is varied, reducing the parameter search to a single dimension. To identify the optimal values of $a$ that maximize $r$, an exhaustive search over $a$ is conducted for different connectivity degrees, with the resulting optimal ratios summarized in Table~\ref{tab:best_ratio_degree}. A rigorous analysis of the energy spectrum of the $n$-qubit Hamiltonian cost function, including both ground-state and excited-state energy levels, is presented in Appendix~\ref{appen_C}.

\begin{table}[htbp]
\centering
\caption{Optimal ratios corresponding to different connectivity degrees. The symbol ``$-$'' indicates that the result is independent of $a$ when the connectivity degree is 0.}
\label{tab:best_ratio_degree}
\begin{tabular}{c|cccccccc}
\hline
Connectivity degree & 0 & 1 & 2 & 3 & 4 & 5 & 6 & 7 \\
\hline
$a$ & $-$ & 1.00 & 1.00 & 2.00 & 1.74 & 1.40 & 1.16 & 1.00 \\
$r$ & 0.1250 & 0.2500 & 0.2727 & 0.2917 & 0.3076 & 0.3390 & 0.3528 & 0.4000 \\
\hline
\end{tabular}
\end{table}

\begin{figure}[h]
\centering
	\includegraphics[width=0.5\linewidth]{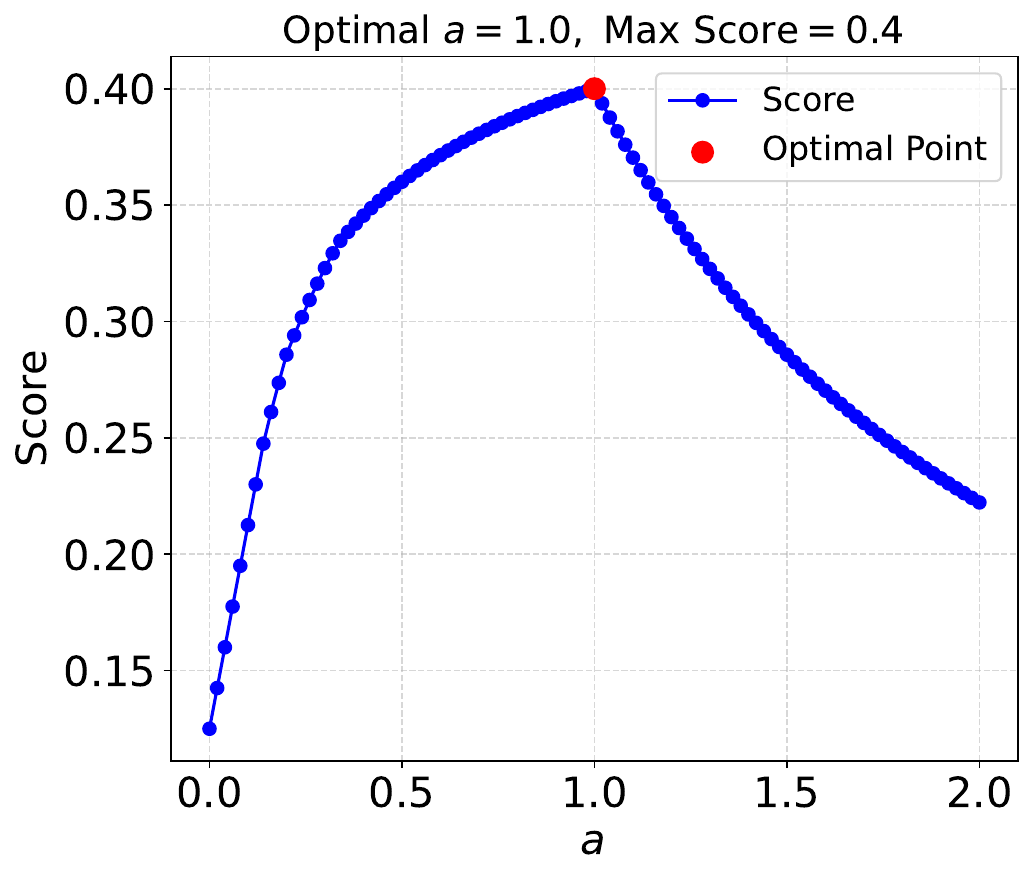}
	\caption{The dependence of score on parameter $a$. The optimal configuration is achieved at $a = 1.0$ with a maximum score of $0.4$ (marked by the red dot).}
\label{fig:ratio_distribution_7}
\end{figure}

Simulation results show that when a 7-regular graph is chosen as the generating graph for the loss function and the absolute values of $w$ and $t$ are both set to $1$, the ratio reaches its maximum value of $0.4$. In this case, the minimum value of the Hamiltonian is $-36$, the second minimum is $-20$, and the maximum value is $4$. To study the influence of different connectivity degrees, the 0-regular graph is selected as a baseline for comparison. For this case, the minimum value of the Hamiltonian is $-8$, the second minimum is $-6$, and the maximum value is $8$. Notably, the 0-regular graph corresponds to the Hamming distance. Based on these results, all subsequent simulations and real-device experiments for S-DES are performed with these fixed parameters.
Figure~\ref{fig:ratio_distribution_7} presents the ratio distribution for the 7-regular case.

The choice of cost function has a profound impact on optimization dynamics and measurement overhead. As such, cost function construction constitutes a central design decision in variational quantum attacks and is a key focus of the benchmarking analysis presented later in this work.

\subsection{Classical Optimization Strategy}

Classical optimization constitutes an essential component of variational quantum attacks, as it governs how information extracted from quantum measurements is used to iteratively update circuit parameters. Within the variational framework, the quantum processor serves primarily as a state preparation and measurement engine, while the classical optimizer navigates the parameter space to minimize the chosen cost function.

Formally, at each iteration $t$, the classical optimizer updates the circuit parameters according to
\begin{equation}
\boldsymbol{\theta}_{t+1} = \mathcal{O}\!\left(\boldsymbol{\theta}_{t},\, \hat{C}(\boldsymbol{\theta}_{t})\right),
\end{equation}
where $\hat{C}(\boldsymbol{\theta}_{t})$ denotes an estimate of the cost function obtained from finite-shot quantum measurements, and $\mathcal{O}(\cdot)$ represents a classical update rule. The stochastic nature of $\hat{C}$, arising from measurement noise and finite sampling, distinguishes variational quantum attacks from deterministic classical optimization problems.

In the context of cryptanalytic tasks, the optimization landscape is typically highly non-convex and may exhibit flat regions or sharp local minima. As a result, the choice of classical optimization strategy significantly influences convergence speed, stability, and robustness. Gradient-based methods~\cite{daoud2023gradient} can exploit smooth cost landscapes and enable rapid convergence when reliable gradient estimates are available. However, they may suffer from noise amplification and barren plateau effects in high-dimensional parameter spaces.
Gradient-free optimization strategies, such as simplex-based~\cite{nelder1965simplex} or evolutionary algorithms~\cite{back1993overview}, offer an alternative that is often more resilient to measurement noise and non-smooth cost functions. While gradient-free methods may require a larger number of function evaluations, their robustness to noisy and non-smooth objective landscapes makes them particularly suitable for small-scale cryptographic cost functions arising from discrete or combinatorial structures in the NISQ era.

At the framework level, we treat the classical optimizer as a modular component that interfaces with the quantum circuit through measured cost values only. This abstraction allows different optimization strategies to be substituted without modifying the underlying quantum circuit design or cost function definition. In later sections, we benchmark representative gradient-based and gradient-free (Nelder-Mead method) optimizers to evaluate their impact on convergence behavior and effective time complexity.

Overall, classical optimization plays a dual role in variational quantum attacks: it enables practical implementation under NISQ constraints and provides a flexible mechanism for incorporating problem-specific heuristics into the attack process. Understanding and selecting appropriate optimization strategies is therefore essential for achieving efficient and reliable variational quantum cryptanalysis.

\section{Benchmarking Results}

In this section, we report the benchmarking results of variational quantum attack algorithms applied to S-DES.
Building upon the modular framework introduced in Sections~\ref{sec_II} and \ref{sec_III}, we conduct a systematic and controlled
evaluation of different combinations of initialization strategies, Ansatz designs, cost functions, and classical
optimization strategies using large-scale classical simulations.
Rather than focusing on a single optimized configuration, our objective is to characterize how these modular
design choices jointly influence the performance of variational attacks.
Specifically, we assess convergence behavior, success probability, and effective time complexity under unified
simulation settings, enabling a fair and quantitative comparison across different algorithmic configurations.

\subsection{Simulation Setup}

All benchmarking results are obtained through classical simulations of variational quantum circuits. The target cryptography is S-DES, which preserves the essential structural features of standard block cryptography while allowing exhaustive simulation of the full key space. The secret key is encoded into a quantum register whose size matches the key length of S-DES, with additional ancillary qubits introduced when required by the Ansatz or cost function construction.

For each experiment, the variational quantum circuit prepares a parameterized quantum state over the key register. Measurement outcomes are sampled to estimate the value of the chosen cost function, which is then fed into a classical optimization loop. To ensure fair comparison across different configurations, we fix number of measurement shots per iteration, and stopping criteria whenever applicable. Classical optimizers are initialized with identical hyperparameter settings across comparable experiments.

30 independent runs are performed for each configuration in order to capture statistical variations in convergence behavior. All reported results are averaged over these runs, and the corresponding success probabilities are estimated based on the fraction of trials that successfully recover the correct secret key within the predefined iteration budget.

We set the convergence threshold of the loss function to $-20$ for the 7-regular graph-structured loss function and to $-6$ for the 0-regular graph-structured loss function.
Extensive simulation results indicate that improper choices of the restart thresholds, whether too large or too small, significantly increase the overall attack complexity. When the threshold is set too large, small fluctuations of the loss function near the initial point may prematurely satisfy the restart condition, leading to frequent restarts and causing the optimization trajectory to oscillate around the initial region. Conversely, if the threshold is set too small, the optimizer may remain trapped in a local minimum for many ineffective iterations before the restart condition is triggered, resulting in unnecessary computational overhead.To balance convergence efficiency and global exploration, we adopt a piecewise restart-threshold strategy in the optimization process based on the graph-structured loss function and Hamming distance loss function, with the thresholds dynamically adjusted according to the optimization stage.

For the 7-regular graph-structured loss function, the restart criterion is defined as

\begin{equation}
\lvert \mathrm{cost}_0 - \mathrm{cost}_{-1} \rvert
<
\begin{cases}
1.00, & \mathrm{cost}_0 < -13, \\
0.33, & \mathrm{cost}_0 < -7, \\
0.17, & \text{otherwise}.
\end{cases}
\end{equation}

where $\text{cost}_0$ denotes the current value of the cost function, and $\text{cost}_{-i}$ denotes the $i$-th previous value of the cost function. The optimization is automatically restarted with a newly randomized set of parameters. This mechanism effectively prevents the optimizer from becoming trapped in local minima.

The restart criterion is defined to scale proportionally with the width of the cost landscape. For the $0$-regular graph-structured loss function, the restart criterion is defined as follows:
\begin{equation}
\lvert \mathrm{cost}_0 - \mathrm{cost}_{-1} \rvert
<
\begin{cases}
0.40, & \mathrm{cost}_0 < -5.2, \\
0.14, & \mathrm{cost}_0 < -2.8, \\
0.06, & \text{otherwise}.
\end{cases}
\label{eq:convergence_criterion}
\end{equation}

\begin{figure}[h]
\centering
	\includegraphics[width=0.5\linewidth]{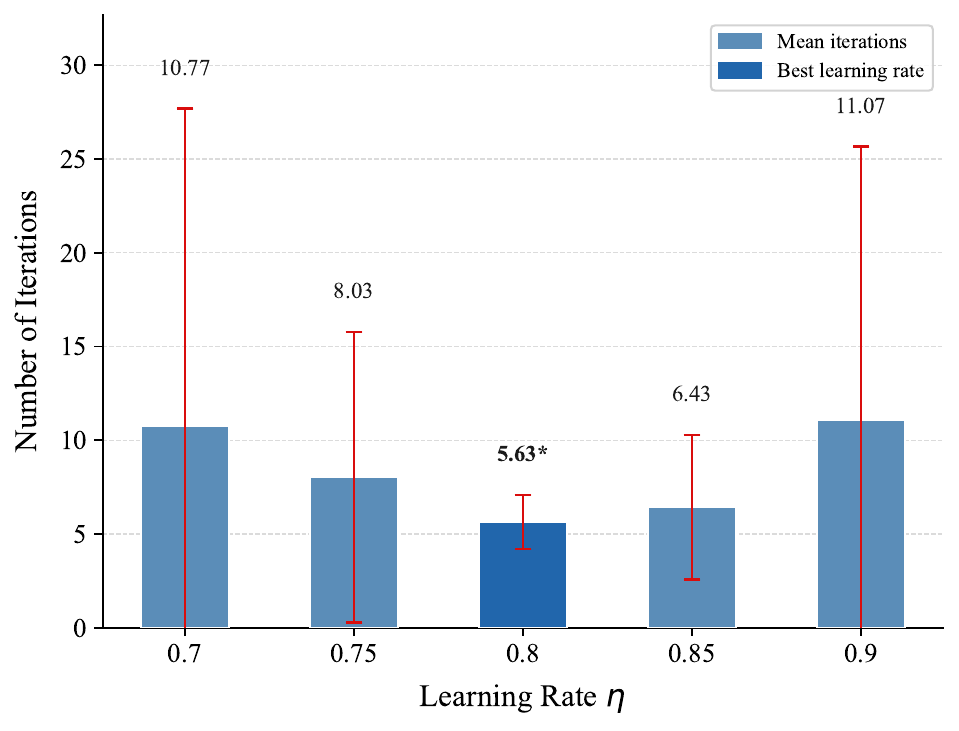}
	\caption{Complexity comparison under different learning rates: Each bar represents the average iteration count over 30 independent runs at the learning rates 0.7, 0.75, 0.8, 0.85, and 0.9, with numerical values indicating the mean complexity. Error bars denote the standard deviation. Note that for learning rates $\eta = 0.7$ and $\eta = 0.9$, the lower portions of the error bars are truncated as they extend below the x-axis; this clipping is applied for visual clarity and does not affect the reported mean values.}
\label{fig:comparision_different_types}
\end{figure}

To determine an appropriate learning rate for the gradient descent method, we conduct a systematic parameter-sweep study. As an illustrative example, we adopt a uniform superposition as the initial state, a non-unitary Ansatz, a 7-regular graph-structured cost function, and a gradient-descent-based classical optimization strategy. The corresponding results are shown in Fig.~\ref{fig:comparision_different_types}, where the number of optimization iterations is reported for different learning rates, providing quantitative guidance for hyperparameter selection.

As observed from Fig.~\ref{fig:comparision_different_types}, the attack performance is optimal when the learning rate is set to $0.8$. Under this hyperparameter configuration, simulations conducted on 30 randomly selected plaintext–ciphertext pairs yield the highest attack success probability. 
These results demonstrate that the chosen hyperparameter configuration provides both high convergence efficiency and robust optimization behavior. The results for the other combinations are provided in Appendix~\ref{appen_D}.

\subsection{Convergence and Success Probability}

We first analyze the convergence behavior of different variational quantum attack configurations. Convergence is quantified by the evolution of the cost function value as a function of optimization iterations, as well as by the probability of measuring the correct key state from the final variational quantum state.

Across all tested configurations, the choice of initialization strategy has a pronounced impact on early-stage convergence. Uniform superposition initialization generally leads to stable but relatively slow convergence, reflecting the lack of prior bias toward the correct key. In contrast, Grover-enhanced initialization consistently accelerates convergence by increasing the initial overlap between the variational state and the target key state.

\begin{figure}[h]
\centering
\begin{minipage}[c]{0.48\linewidth}
    \centering
    \includegraphics[width=\linewidth]{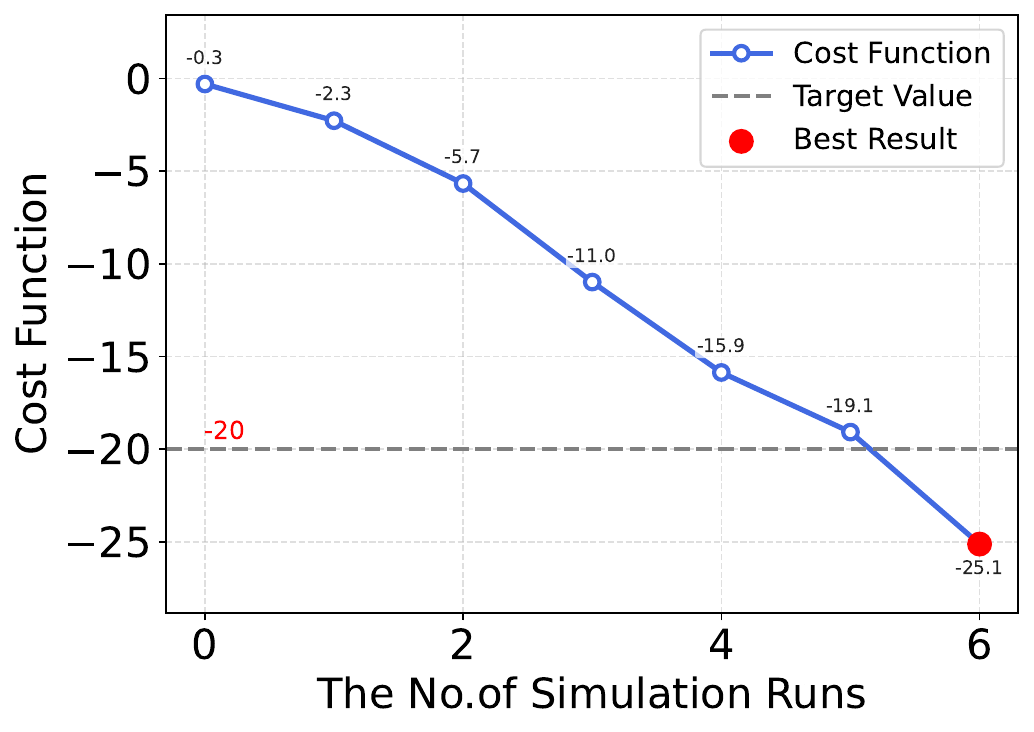}
    
\end{minipage}
\hfill
\begin{minipage}[c]{0.48\linewidth}
    \centering
    \includegraphics[width=\linewidth]{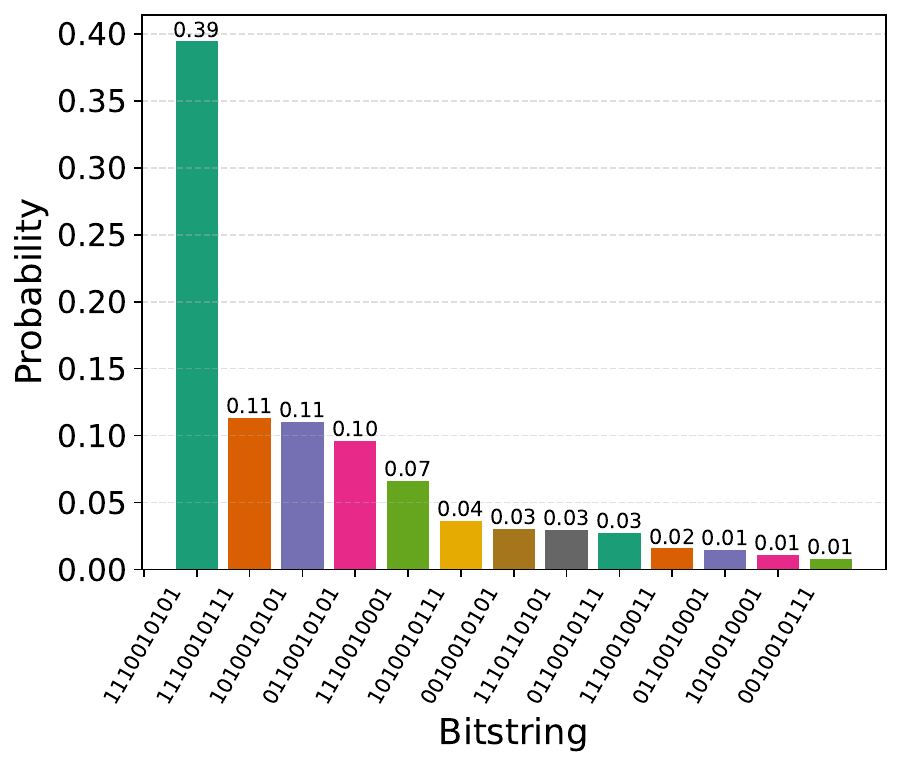}
    
\end{minipage}

\vspace{0.4em}

% ---------- Labels ----------
\begin{minipage}{0.48\textwidth}
\centering
\textbf{(a)}
\end{minipage}
\hfill
\begin{minipage}{0.48\textwidth}
\centering
\textbf{(b)}
\end{minipage}
\caption{Optimization results.
(a) Evolution of the loss function as a function of the iteration number under the optimal hyperparameter configuration. The horizontal dashed line at $-20$ denotes the convergence threshold of the loss function.
(b) Probability distribution of the final quantum state. The x-axis shows the corresponding bit strings. The first bar indicates the probability of the target state, and the remaining bars are sorted in descending order of probability.
}
\label{fig:convergency_distribution}
\end{figure}

The structure of the Ansatz plays a crucial role in determining convergence stability. Shallow unitary Ansatz exhibit robust performance across different runs, albeit with limited expressive power, whereas deeper or non-unitary-inspired Ansatz tend to converge more rapidly at the expense of increased runtime per iteration. These results reveal a fundamental trade-off between expressibility and trainability that is characteristic of variational quantum algorithms. Figure~\ref{fig:convergency_distribution}(a) presents a representative optimization trajectory obtained under the optimal hyperparameter configuration using the same modular setting as in Fig.~\ref{fig:comparision_different_types}. The loss function decreases monotonically with the iteration number and reaches the target solution within 6 iterations, indicating stable and efficient convergence.

The success probability is quantified by the frequency with which the correct key is obtained with high probability at the end of the optimization process. Figure~\ref{fig:convergency_distribution}(b) depicts the corresponding probability distribution of the final quantum state.
Through computation, we identify eight candidate keys corresponding to the known plaintext--ciphertext pair, namely \texttt{0110000011}, \texttt{0110010111}, \texttt{0111001011}, \texttt{0111011111}, \texttt{1010000001}, \texttt{1010010101}, \texttt{1110000001}, and \texttt{1110010101}. Among these, the occurrence probabilities of \texttt{0110010111}, \texttt{1010010101}, and \texttt{1110010101} are $0.03$, $0.11$, and $0.39$, respectively, while those of the remaining candidates are negligible. Therefore, the overall success probability, defined as the sum of these probabilities, is approximately $0.53$.

\subsection{Effective Time Complexity Comparison}

Beyond convergence behavior, we compare different attack configurations in terms of effective time complexity. Rather than focusing solely on asymptotic query complexity, we define effective time complexity as the total computational effort required to achieve successful key recovery with high probability. This metric incorporates the number of optimization iterations, and the measurement cost per iteration, and the circuit depth associated with each Ansatz. For example, a single Grover iteration is counted as one optimization iteration.

Our benchmarking results indicate that carefully designed variational quantum attacks can substantially reduce the effective time complexity compared with naive exhaustive search strategies. Table~\ref{tab:iteration_comparison} presents a comparison of the effective time complexity across several optimization configurations. Overall, the results show that both the choice of ansatz and the structure of the cost function have a significant impact on the convergence efficiency.

For the $7$-regular cost function optimized with gradient descent and the $0$-regular cost function optimized with the Nelder--Mead method, the Grover-enhanced initialization achieves the best performance, requiring the smallest average number of iterations to converge. This suggests that, when combined with a well-structured cost landscape, increasing the initial overlap with the target state can indeed accelerate convergence. In addition, under the same setting, the non-unitary ansatz outperforms the unitary ansatz, indicating enhanced expressibility and greater optimization flexibility.
By comparing the performance of different optimizers, namely gradient descent and Nelder--Mead, we further observe that gradient descent generally achieves superior performance in the considered configurations.

\begin{table}[H]
\centering
\caption{Average number of optimization iterations under different modular configurations}
\label{tab:iteration_comparison}
\begin{tabular}{l l l l c}
\toprule
\textbf{Initialization} & \textbf{Ansatz} & \textbf{Cost Function} & \textbf{Optimizer} & \textbf{Avg. Iter.} \\
\midrule
Uniform superposition & Unitary & 7-regular & Gradient descent & 7.70 \\
Uniform superposition & Non-unitary & 7-regular & Gradient descent & 5.63 \\
Grover-enhanced & Unitary & 7-regular & Gradient descent & 5.33 \\
Uniform superposition & Unitary & 0-regular & Gradient descent & 14.93 \\
Uniform superposition & Unitary & 0-regular & Nelder--Mead & 291.14 \\
Grover-enhanced & Unitary & 0-regular & Nelder--Mead & 266.02 \\
\bottomrule
\end{tabular}
\end{table}

When compared with Grover-based quantum search, variational quantum attacks do not offer an asymptotic advantage in the worst case. However, under realistic NISQ constraints—where circuit depth, coherence time, and measurement overhead are limited—the variational approach demonstrates favorable trade-offs. By shifting part of the computational burden to classical optimization and exploiting problem-specific structure, variational quantum attacks can achieve lower effective runtime within feasible hardware limits.

These results suggest that variational quantum attack algorithms constitute a practical intermediate strategy between classical brute-force search and fully fault-tolerant quantum search. The modular benchmarking framework presented here provides a quantitative basis for evaluating such trade-offs and for guiding the design of future quantum cryptanalytic implementations.

\subsection{Unified Performance Metric}

While convergence behavior and success probability provide essential insights into the effectiveness of a given variational quantum attack configuration, a fair comparison across different modular combinations requires a unified metric that accounts for both algorithmic performance and resource consumption. In particular, circuit depth, qubit overhead, and the number of optimization iterations all contribute significantly to the practical feasibility of an attack.

To this end, we introduce an integrated performance metric that combines these factors into a single figure of merit. For a given configuration, the metric is defined as
\begin{equation}
\mathcal{M}
=
\frac{P_{\mathrm{succ}}}
{\alpha\, D
+\beta\, Q
+\gamma\, N_{\mathrm{eva}}},
\label{eq:unified_metric}
\end{equation}
where $P_{\mathrm{succ}} = M/N$ denotes the success probability of recovering the correct key. Here, $N$ is the total number of simulation runs, and $M$ is the number of runs in which the key is successfully extracted within a preset iteration threshold. Furthermore, $D$ denotes the two-qubit gate circuit depth of the corresponding Ansatz, $Q$ is the total number of qubits used in the circuit, and $N_{\mathrm{eva}}$ represents the number of classical optimization evaluations required for the loss function to fall below a predefined threshold. For example, in Combination~I, each iteration requires $11$ function evaluations: $10$ evaluations for updating the $10$ variational parameters and one additional evaluation to compute the loss function after the parameter update. The coefficients $\alpha$, $\beta$, and $\gamma$ are positive weighting factors that allow different resource costs to be emphasized depending on the target hardware or application scenario.

This formulation reflects the intuition that a favorable variational attack should not only achieve a high success probability, but should also do so with minimal quantum and classical resources. In particular, configurations with deeper circuits or larger qubit requirements are penalized, as they are more susceptible to noise and incur higher implementation overhead on near-term quantum devices.

Using Eq.~\eqref{eq:unified_metric}, we evaluate and compare different combinations of initialization strategies, Ansatz structures, cost functions, and classical optimization methods on an equal footing. The data used for this evaluation are summarized in Table~\ref{tab:data_comparison}, with combination indices corresponding to the configurations listed in Table~\ref{tab:iteration_comparison}. The resulting metric enables a systematic ranking of configurations and highlights those that achieve the best balance between expressibility, trainability, and resource efficiency.
\begin{table}[H]
\centering
\caption{Summary of performance-related metrics for different modular combinations}
\label{tab:data_comparison}
\setlength{\tabcolsep}{10pt}
\renewcommand{\arraystretch}{1.15}
\begin{tabular}{c c c c c c}
\toprule
\textbf{Combination} 
& $P_{\mathrm{succ}}$ 
& $D$ 
& $Q$ 
& $N_{\mathrm{eva}}$ 
& $\mathcal{M}$\\
\midrule
I   & 1 & 10 & 10 & 84.70 & $\frac{1}
{10\, \alpha 
+10\, \beta
+84.70\, \gamma}$\\
II  &  1   & 19  & 19   & 112.67 & $\frac{1}
{19\, \alpha 
+19\, \beta
+112.67\, \gamma}$ \\
III &   1  &  $\mathcal{O}(10^3)$   &  18  & 58.67 & $\frac{1}
{\mathcal{O}(10^3)\, \alpha 
+18\, \beta
+58.67\, \gamma}$\\
IV  &  0.97   &  10  &  10  & 164.27 & $\frac{0.97}
{10\, \alpha 
+10\, \beta
+164.27\, \gamma}$ \\
V   &  0.72   &  10  &  10  & 426.69 & $\frac{0.72}
{10\, \alpha 
+10\, \beta
+426.69\, \gamma}$\\
VI  &  0.76   &  $\mathcal{O}(10^3)$  &  18  & 391.17 & $\frac{0.76}
{\mathcal{O}(10^3)\, \alpha 
+18\, \beta
+391.17\, \gamma}$ \\
\bottomrule
\end{tabular}
\end{table}

Based on the data reported in Table~\ref{tab:data_comparison}, several qualitative trends can be observed. 
Configurations with shallow unitary Ansatze and modest qubit requirements, such as Combinations~I, IV, and V, achieve relatively high success probabilities while maintaining low circuit depth and qubit overhead. 
As a result, these configurations are expected to score favorably under the integrated metric~$\mathcal{M}$, reflecting a balanced trade-off between algorithmic performance and resource efficiency.
In contrast, configurations involving deeper circuits, exemplified by Combinations~III and VI, exhibit substantially larger two-qubit gate depths on the order of $\mathcal{O}(10^3)$. 
Although such configurations can still attain moderate success probabilities, the significantly increased circuit depth and optimization cost penalize their overall metric value, highlighting their reduced practicality on near-term quantum hardware. 
This observation underscores the importance of limiting circuit depth when evaluating variational quantum attacks in the NISQ regime.
Furthermore, while higher success probability generally correlates with better performance, the table demonstrates that success probability alone is insufficient as a sole evaluation criterion. 
For instance, Combination~II achieves a high success probability but requires increased qubit resources and a larger number of optimization iterations, which diminishes its overall efficiency when resource consumption is taken into account.

Overall, the comparison confirms that the proposed unified metric captures essential trade-offs that are not visible from convergence behavior or success probability alone. 
By jointly accounting for quantum resources and classical optimization cost, the metric provides a more realistic basis for ranking variational quantum attack configurations and identifying those that are most suitable for implementation on near-term quantum devices.

\section{Discussion and Outlook}

The benchmarking results presented in this work provide several insights into the design principles and practical performance of variational quantum attack algorithms on symmetric cryptosystems. Rather than identifying a single universally optimal configuration, our analysis highlights how different algorithmic modules interact and how their interplay determines overall efficiency under realistic resource constraints.

A central observation concerns the role of initialization strategies in shaping the optimization landscape. While Grover-enhanced initialization achieves the lowest iteration count among all tested configurations (Table~\ref{tab:iteration_comparison}), this advantage comes at the cost of significantly increased circuit depth. As shown in Table~\ref{tab:data_comparison}, the Grover-enhanced configuration requires a two-qubit gate depth on the order of $\mathcal{O}(10^3)$, which renders it impractical under typical NISQ constraints where noise accumulation scales with circuit depth. In contrast, the uniform superposition initialization (Combination~I) achieves comparable success probability with only $10$ two-qubit gates, demonstrating that simpler initialization strategies can offer superior resource-efficiency trade-offs on near-term hardware. This finding underscores the importance of jointly evaluating both algorithmic performance and quantum resource consumption when selecting initialization strategies for variational quantum attacks.

The choice of Ansatz further reveals a fundamental trade-off between expressibility and trainability~\cite{sim2019expressibility}. While deeper or non-unitary-inspired Ansatze can, in principle, accelerate convergence toward the correct key, our results indicate that shallow unitary Ansatze often achieve comparable or superior overall performance when circuit depth is constrained. The non-unitary Ansatz (Combination~II) requires approximately twice the circuit depth of its unitary counterpart (Combination~I) while delivering similar success probabilities. These findings suggest that, for S-DES-scale problems, the expressibility gains of deeper Ansatze do not justify their additional resource overhead. In the NISQ regime, robustness and shallow circuit depth are often more critical than expressibility.

Cost function design emerges as another decisive factor in determining the practical viability of variational quantum attacks. As shown in Table~\ref{tab:data_comparison}, configurations based on 7-regular graph-structured cost functions (Combinations~I, II, and III) all achieve $P_{\mathrm{succ}} = 1.0$, whereas configurations based on 0-regular Hamming-distance cost functions (Combinations~IV, V, and VI) yield substantially lower success probabilities ($0.97$, $0.72$, and $0.76$, respectively). This performance gap is attributable to the smoother optimization landscape induced by the graph-based Hamiltonian formulation, which facilitates gradient-based classical optimization. The analytical results presented in Appendix~\ref{appen_C} further demonstrate that the normalized spectral gap $r$ of the $n$-qubit $d$-regular Ising Hamiltonian increases monotonically with the connectivity degree $d$, providing a principled theoretical foundation for the empirical superiority of higher-connectivity cost functions. These findings underscore that cost function selection, particularly the choice of graph connectivity, is a key design lever for achieving reliable key recovery in variational quantum cryptanalysis.

From a broader perspective, our results clarify the relationship between variational quantum attacks and Grover-based quantum search. While variational approaches do not surpass Grover's algorithm in asymptotic query complexity, the modular benchmarking framework developed in this work demonstrates that carefully designed variational configurations can achieve favorable effective time-complexity trade-offs under realistic NISQ constraints. In particular, the proposed unified metric $\mathcal{M}$ reveals that shallow unitary configurations (Combinations~I, IV, and V) offer superior resource-efficiency trade-offs compared to deep-circuit alternatives (Combinations~III and VI), which, despite achieving competitive iteration counts, incur substantially larger circuit depth overhead. This integrated perspective highlights trade-offs that are not captured by asymptotic analyses or single-metric evaluations alone, and underscores the practical value of modular variational design in near-term quantum cryptanalysis.

Overall, the modular benchmarking framework developed in this work provides a unified and reproducible methodology for evaluating variational quantum attacks beyond isolated case studies. The key insight is that no single modular choice dominates across all resource regimes; instead, optimal configurations must be selected based on the specific constraints of the target hardware. Looking ahead, this framework can be extended to reduced-round or lightweight variants of AES, as well as to symmetric primitives in the post-quantum setting. Incorporating explicit noise models and hardware-specific constraints will be essential for assessing performance on real quantum devices. More broadly, the modular perspective adopted here may serve as a foundation for hybrid fault-tolerant--variational cryptanalytic strategies in future quantum architectures.

\section{Data and Code Availability}
The source code used in this work is publicly available at \texttt{[https://gitee.com/spaceofwang/benchmark]}. The codebase includes all modular configurations and the classical simulation framework used in our experiments.

\section{Acknowledgements}
Z.W. acknowledges support from the Research Initiation Fund of the Quantum Science Center of the Guangdong–Hong Kong–Macao Greater Bay Area (Grant No. QD2305001). K.W. was also supported by this fund. 
Z.W. further acknowledges support from the National Natural Science Foundation of China (Grant No. 62501522) and the Guangdong Provincial Quantum Science Strategic Initiative (Grant No. GDZX2503005). 
S.W. acknowledges support from the Beijing Nova Program (Grant Nos. 20230484345 and 20240484609). 
This work was also supported by the National Natural Science Foundation of China (Grant Nos. 62471046 and 62571050).

\bibliography{SDES}

\clearpage

\appendix
\onecolumngrid

\renewcommand{\thefigure}{A.\arabic{figure}}
\renewcommand{\thetable}{A.\arabic{table}}
\renewcommand{\theequation}{A.\arabic{equation}}
\setcounter{figure}{0} % 重置图表编号
\setcounter{table}{0} % 重置表格编号
\setcounter{equation}{0}

\section{The details of S-DES}\label{appen_A}

\noindent\textbf{Sub-key generation.}

The initial 10-bit key is denoted as
\(
(k_1,k_2,k_3,k_4,k_5,k_6,k_7,k_8,k_9,k_{10})
\).
Two 8-bit sub-keys, $K_1$ and $K_2$, are generated according to
\begin{equation}
\begin{aligned}
K_{1} &= \mathrm{P8}\!\left[\mathrm{Shift}_1\!\left(\mathrm{P10}(\mathrm{key})\right)\right], \\
K_{2} &= \mathrm{P8}\!\left[\mathrm{Shift}_2\!\left(\mathrm{Shift}_1\!\left(\mathrm{P10}(\mathrm{key})\right)\right)\right].
\end{aligned}
\label{eq:gkey}
\end{equation}

The permutation function $\mathrm{P10}$ is defined as
\begin{equation}
\begin{aligned}
\mathrm{P10}(k_1,\ldots,k_{10})
=
(k_3, k_5, k_2, k_7, k_4, k_{10}, k_1, k_9, k_8, k_6).
\end{aligned}
\label{eq:p10}
\end{equation}
That is, each output bit is selected from the corresponding input position specified in Eq.~\eqref{eq:p10}.

After applying $\mathrm{P10}$, the resulting 10-bit string is divided into two 5-bit halves, each of which is circularly left-shifted by one bit, denoted as $\mathrm{Shift}_1$.  
The sub-key $K_1$ is then obtained by selecting 8 bits via the permutation $\mathrm{P8}$:
\begin{equation}
\begin{array}{cccccccc}
\hline \multicolumn{8}{c}{\mathrm{P8}} \\
\hline
6 & 3 & 7 & 4 & 8 & 5 & 10 & 9 \\
\hline
\end{array}.
\label{eq:p8}
\end{equation}
Here, the indices refer to the positions in the 10-bit string produced by $\mathrm{P10}$.

To generate $K_2$, the two 5-bit halves produced by $\mathrm{Shift}_1$ are further circularly left-shifted by two bits, denoted as $\mathrm{Shift}_2$, followed by the same $\mathrm{P8}$ permutation defined in Eq.~\eqref{eq:p8}.

\medskip
\noindent\textbf{Encryption procedure.}

Given an 8-bit plaintext, the encryption starts with the initial permutation $\mathrm{IP}$,
\begin{equation}
\begin{array}{cccccccc}
\hline \multicolumn{8}{c}{\mathrm{IP}} \\
\hline
2 & 6 & 3 & 1 & 4 & 8 & 5 & 7 \\
\hline
\end{array},
\end{equation}
which preserves all information while rearranging the bit order.  
The inverse permutation $\mathrm{IP}^{-1}$ is given by
\begin{equation}
\begin{array}{cccccccc}
\hline \multicolumn{8}{c}{\mathrm{IP}^{-1}} \\
\hline
4 & 1 & 3 & 5 & 7 & 2 & 8 & 6 \\
\hline
\end{array}.
\end{equation}

The core component of S-DES is the round function $f_K$, defined as
\begin{equation}
f_K(L,R) = \bigl(L \oplus F(R,K),\, R\bigr),
\label{eq:fk}
\end{equation}
where $L$ and $R$ denote the left and right 4-bit halves of the input, respectively, $K$ is the sub-key, and $\oplus$ represents the XOR operation.

The function $F$ maps a 4-bit input $(n_1,n_2,n_3,n_4)$ to a 4-bit output.  
It begins with an expansion-permutation $\mathrm{E/P}$:
\begin{equation}
\begin{array}{cccccccc}
\hline \multicolumn{8}{c}{\mathrm{E/P}} \\
\hline
4 & 1 & 2 & 3 & 2 & 3 & 4 & 1 \\
\hline
\end{array},
\end{equation}
which produces
\begin{equation}
\begin{array}{l|ll|l}
n_4 & n_1 & n_2 & n_3 \\
n_2 & n_3 & n_4 & n_1
\end{array}.
\end{equation}

After XORing with the 8-bit sub-key
\(K=(k_{11},k_{12},\ldots,k_{18})\),
the result is written as
\begin{equation}
\begin{array}{l|ll|l}
p_{0,0} & p_{0,1} & p_{0,2} & p_{0,3} \\
p_{1,0} & p_{1,1} & p_{1,2} & p_{1,3}
\end{array}.
\label{eq:ff}
\end{equation}

The first and second rows of Eq.~\eqref{eq:ff} are fed into the S-boxes $S_0$ and $S_1$, respectively, each producing a 2-bit output.  
The S-boxes are defined as
\begin{equation}
\mathrm{S}_0=
\bordermatrix{
& 0 & 1 & 2 & 3 \cr
0 & 1 & 0 & 3 & 2 \cr
1 & 3 & 2 & 1 & 0 \cr
2 & 0 & 2 & 1 & 3 \cr
3 & 3 & 1 & 3 & 2
},
\hspace{0.5cm}
\mathrm{S}_1=
\bordermatrix{
& 0 & 1 & 2 & 3 \cr
0 & 0 & 1 & 2 & 3 \cr
1 & 2 & 0 & 1 & 3 \cr
2 & 3 & 0 & 1 & 0 \cr
3 & 2 & 1 & 0 & 3
}.
\end{equation}

For each S-box, the first and fourth bits of the 4-bit input specify the row index, while the second and third bits specify the column index. The corresponding entry yields a 2-bit output.

The resulting 4-bit string is finally permuted by $\mathrm{P4}$,
\begin{equation}
\begin{array}{cccc}
\hline \multicolumn{4}{c}{\mathrm{P4}} \\
\hline
2 & 4 & 3 & 1 \\
\hline
\end{array},
\end{equation}
which constitutes the output of the function $F$.

The round function $f_K$ modifies only the left 4 bits. Between the two rounds, the swap operation $\mathrm{SW}$ exchanges the left and right halves, ensuring that different bit positions are processed in the second round. 

\renewcommand{\thefigure}{B.\arabic{figure}}
\renewcommand{\thetable}{B.\arabic{table}}
\renewcommand{\theequation}{B.\arabic{equation}}
\setcounter{figure}{0} % 重置图表编号
\setcounter{table}{0} % 重置表格编号
\setcounter{equation}{0}
\section{Simulation Results for Different Ratios under Combination I}\label{appen_B}

In this section, we present additional simulation results for different values of the ratio under Combination~I. This setting corresponds to the configuration with uniform superposition initialization, a unitary ansatz, a 7-regular graph-structured cost function, and gradient descent optimization. The numerical results are summarized in Table~\ref{tab:ratio_comparison}, where the ratio is defined as the spectral gap between the ground state and the first excited state divided by the width of the energy spectrum of the corresponding Hamiltonian.

\begin{table}[H]
\centering
\caption{Simulation results for different ratios under Combination I}
\label{tab:ratio_comparison}
\renewcommand{\arraystretch}{1.3}
\begin{tabular}{@{\hspace{1em}} c @{\hspace{2em}} c @{\hspace{2em}} c @{\hspace{2em}} c @{\hspace{2em}} c @{\hspace{1em}}}
\toprule
$a$ & \textbf{Ratio} & \textbf{Convergence Threshold} & \textbf{The optimal learning rate} & \textbf{Avg.\ Iter.} \\
\midrule
0.1  & 0.2125 & $-7.4$  & 0.85 & $10.67$ \\
0.2  & 0.2857 & $-8.8$  & 0.9  & $11.20$ \\
0.3  & 0.2857 & $-8.8$  & 0.9  & $7.73$  \\
0.4  & 0.3229 & $-10.2$ & 0.95 & $10.73$ \\
0.5  & 0.3455 & $-11.6$ & 0.9  & $8.63$  \\
0.6  & 0.3600 & $-13.0$ & 0.8  & $10.93$ \\
0.7  & 0.3714 & $-14.4$ & 0.75 & $6.60$  \\
0.8  & 0.3806 & $-15.8$ & 0.75 & $7.43$  \\
0.9  & 0.3946 & $-18.6$ & 0.75 & $7.67$  \\
0.95 & 0.3974 & $-19.3$ & 0.7  & $8.23$  \\
\bottomrule
\end{tabular}
\end{table}

From Table~\ref{tab:ratio_comparison}, one observes that as the parameter $a$ increases, the corresponding ratio exhibits a monotonic growth trend. This indicates that enlarging $a$ effectively improves the relative spectral gap of the Hamiltonian, which is expected to enhance the optimization landscape by increasing the distinguishability between the ground state and low-lying excited states.

\begin{figure}[H]
    \centering
    \includegraphics[width=0.55\linewidth]{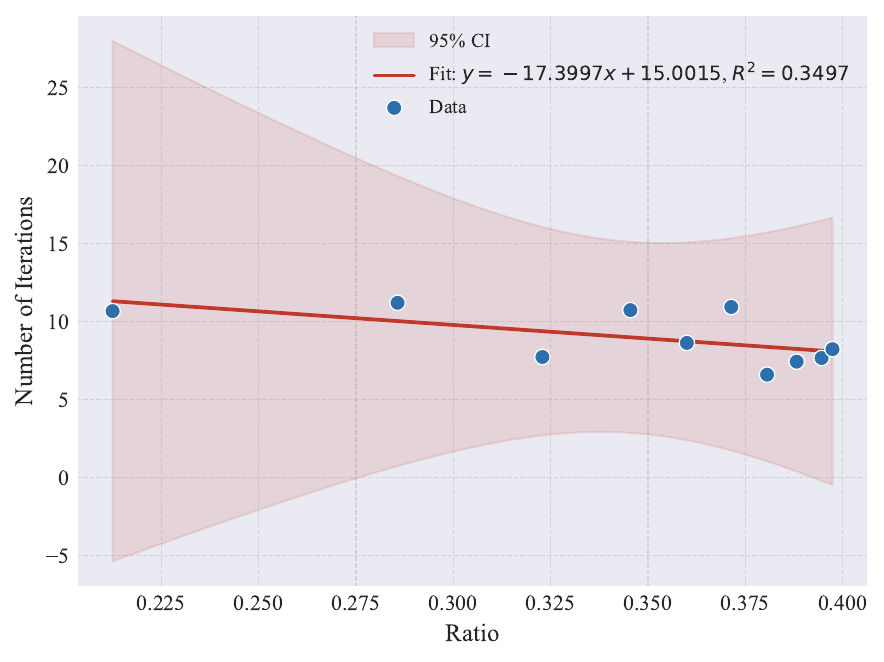}
    \caption{Average number of iterations as a function of the ratio under Combination~I. The solid line represents a linear fit, and the shaded region denotes the 95\% confidence interval.}
    \label{fig:appendix_ratio}
\end{figure}

To further investigate the relationship between the ratio and the optimization performance, we plot in Fig.~\ref{fig:appendix_ratio} the average number of iterations required for convergence as a function of the ratio. A linear regression is performed to quantify this dependence.

As shown in Fig.~\ref{fig:appendix_ratio}, there exists a clear negative correlation between the ratio and the average number of iterations. In particular, the fitted linear model suggests that larger ratios generally lead to fewer iterations required for convergence. This behavior is consistent with the intuition that a larger normalized spectral gap yields a more favorable optimization landscape, thereby accelerating convergence of the gradient-based algorithm.
We also note that although the overall trend is approximately linear, fluctuations are present due to stochastic effects in the optimization process and finite sampling. Nevertheless, the observed correlation provides strong empirical evidence that increasing the ratio is beneficial for improving the efficiency of the variational optimization.

In summary, the results in this section demonstrate that the ratio serves as an effective indicator of optimization performance under Combination~I. A larger ratio not only corresponds to a wider separation between energy levels but also leads to faster convergence in practice.

\renewcommand{\thefigure}{C.\arabic{figure}}
\renewcommand{\thetable}{C.\arabic{table}}
\renewcommand{\theequation}{C.\arabic{equation}}
\setcounter{figure}{0} % 重置图表编号
\setcounter{table}{0} % 重置表格编号
\setcounter{equation}{0}
\section{The Convergence Threshold Determination in VQAAs for $n$-bit Symmetric Cryptosystems}\label{appen_C}
As discussed in Section~III-C and Appendix~B, a larger value of the defined ratio leads to better performance. This naturally raises two important questions: how can this ratio be increased, and what is the corresponding convergence threshold? Addressing these issues is crucial. In this section, we provide a detailed analysis of convergence threshold determination in VQAAs for $n$-bit symmetric cryptosystems.

It is straightforward to show that the energy spectrum of the Hamiltonian
\begin{equation}
H=\sum_{\{i,j\} \in E} w_{ij} Z_i Z_j + \sum_{i=0}^{n-1} t_i Z_i
\end{equation}
coincides with that of the Ising model, which has been extensively studied in statistical physics.

Consider a $d$-regular connected graph $G=(V,E)$ with $|V|=n$ and $|E|=\frac{nd}{2}$. The corresponding Ising Hamiltonian can be written as
\begin{equation}
H = -a \sum_{(i,j) \in E} x_i x_j - b \sum_{i=1}^{n} x_i,
\qquad x_i \in \{-1, +1\},
\end{equation}
where $a > 0$ and $b > 0$. The ground-state energy of this Hamiltonian is non-degenerate and can be expressed as
\begin{equation}
E_0 = -\frac{a n d}{2} - b n.
\end{equation}
This corresponds to the configuration $\boldsymbol{x} = (+1, +1, \dots, +1)$.

To determine the first excited energy, we compare two natural candidate configurations.
The first is the global spin-flip configuration $x_i = -1$ for all $i$, whose energy differs from the ground state by
\begin{equation}
\Delta_{\mathrm{all}} = 2bn.
\end{equation}
The second candidate is obtained by flipping a single vertex relative to the ground state. Since each vertex has degree $d$, this operation modifies $d$ interaction terms, leading to an energy increase
\begin{equation}
\Delta_{\mathrm{one}} = 2da + 2b.
\end{equation}

The first excited energy is therefore given by the minimum of these two possibilities, and the spectral gap $\Delta = E_1 - E_0$ takes the form
\begin{equation}
\Delta =
\begin{cases}
2da + 2b, & a \le \dfrac{b(n-1)}{d}, \\[6pt]
2bn, & a > \dfrac{b(n-1)}{d}.
\end{cases}
\end{equation}
The transition point is determined by the match condition

\begin{equation}
a^* = \frac{b(n-1)}{d},
\end{equation}
at which the two excitation mechanisms become energetically equivalent. In the first case, $\Delta$ is directly proportional to $a$, whereas in the second case it is independent of $a$. Therefore, the optimal value is achieved at the matching point between the two regimes, where $\Delta$ attains its maximum. Consequently, we obtain
\begin{equation}
\Delta_{\max} = \Delta(a^\ast) = 2 d a^\ast + 2 b.
\end{equation}

This result admits a simple interpretation: when the coupling strength $a$ is small, the cost of modifying a limited number of edges is low, and local perturbations dominate. In contrast, for large $a$, edge interactions are dominant, making local flips expensive and rendering the global flip the lowest-energy excitation.

The maximum energy of the Hamiltonian can be expressed in terms of the MaxCut problem on $G$:
In general, computing $\mathrm{MaxCut}(G)$ is NP-hard, which implies that $E_{\max}$ does not admit a closed-form expression for arbitrary $d$-regular graphs. 
However, we can derive an upper bound on $E_{\max}$ by noting that the maximum cut size is at most the total number of edges, i.e., $\mathrm{MaxCut}(G) \le |E| = \frac{nd}{2}$. This leads to the bound $E_{\max} \le \frac{and}{2}$, So the width of the energy spectrum can be bounded as
\begin{equation}
w = E_{\max} - E_0 \le \frac{and}{2} + \frac{and}{2} + bn = and + bn.
\end{equation}
Consequently, the ratio $r$ of the spectral gap to the width can be bounded as
\begin{equation}
r >\frac{2da + 2b}{and + bn}
\end{equation}

Taking the derivative with respect to $d$, we get
\begin{equation}
r' >
\frac{2abn}{(and + bn)^2}.
\end{equation}
This result indicates that $r$ is monotonically increasing with respect to $d$ for $a > 0$ and $b > 0$. Therefore, the convergence threshold can be improved by increasing the degree of the regular graph, which corresponds to introducing more pairwise interactions into the cost function.

Without loss of generality, we set $b = 1$. When the degree reaches its maximum value $d = n - 1$, the matching condition yields $a = 1$. At this point, the convergence threshold for VQAAs is given by
\begin{equation}
\frac{n(3-n)}{2} .
\end{equation}

\renewcommand{\thefigure}{D.\arabic{figure}}
\renewcommand{\thetable}{D.\arabic{table}}
\renewcommand{\theequation}{D.\arabic{equation}}
\setcounter{figure}{0} % 重置图表编号
\setcounter{table}{0} % 重置表格编号
\setcounter{equation}{0}
\section{The simulation results for the different combinations}\label{appen_D}

In Section~IV, we focused on a representative modular configuration to illustrate the hyperparameter selection procedure and the corresponding optimization behavior in detail. Specifically, Fig.~5 presents the hyperparameter landscape used to identify the optimal configuration, and Fig.~6 reports the resulting convergence behavior and final success probability.

In this appendix, we provide complementary simulation results for the remaining five modular combinations discussed in Section~IV-C. For each configuration, we follow the same evaluation protocol as in the main text: we first determine suitable hyperparameters, and then analyze a representative optimization trajectory together with the final probability distribution of the quantum state. These additional results are included here for completeness and to demonstrate the robustness of the observed trends across different algorithmic choices.

\subsection{Combination I}
Combination I corresponds to the configuration with uniform superposition initialization, a unitary Ansatz, a 7-regular graph-structured cost function, and gradient descent optimization, as summarized in Table~II.

\begin{figure}[h]
    \centering
    \includegraphics[width=0.55\linewidth]{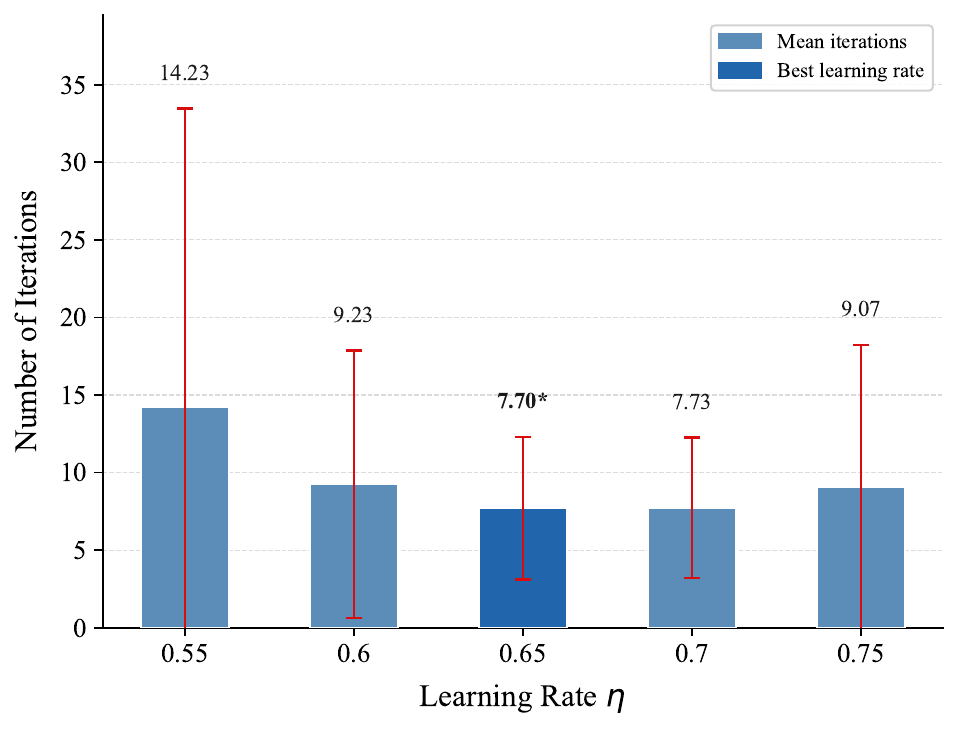}
    \caption{Mean number of optimization iterations as a function of learning rate, evaluated across five learning rate settings ($\eta \in \{0.55, 0.6, 0.65, 0.7, 0.75\}$). Red error bars denote the standard deviation. The optimal learning rate $\eta = 0.65$ (highlighted in dark blue) achieves the lowest mean iteration count of $7.70$, indicating the fastest convergence among all candidates.}

    \label{fig:appendix_comb1_lr}
\end{figure}

Following the same evaluation protocol as in Section~IV, we first determine the optimal learning rate by comparing the average number of optimization iterations under different learning-rate settings. Figure~\ref{fig:appendix_comb1_lr} shows the dependence of the iteration count on the learning rate for this configuration. The optimal learning rate is selected as the one that minimizes the required number of iterations.

\begin{figure}[H]
    \centering
    \begin{minipage}[c]{0.48\linewidth}
        \centering
        \includegraphics[width=\linewidth]{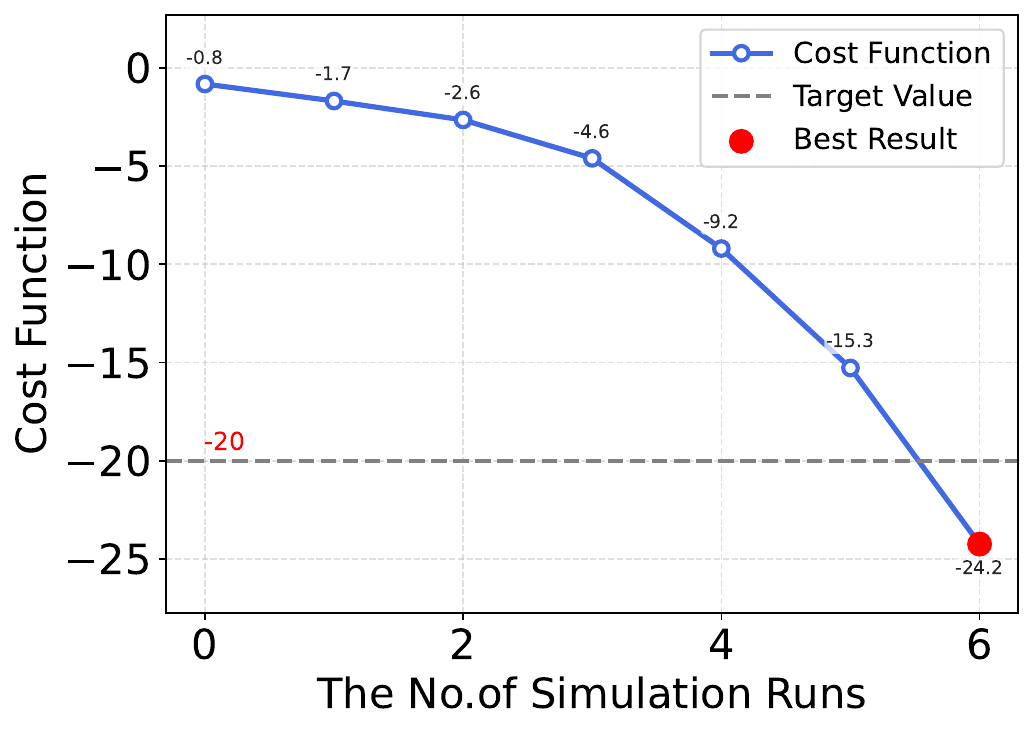}
        \par\vspace{0.5ex}
        (a)
    \end{minipage}\hfill
    \begin{minipage}[c]{0.48\linewidth}
        \centering
        \includegraphics[width=\linewidth]{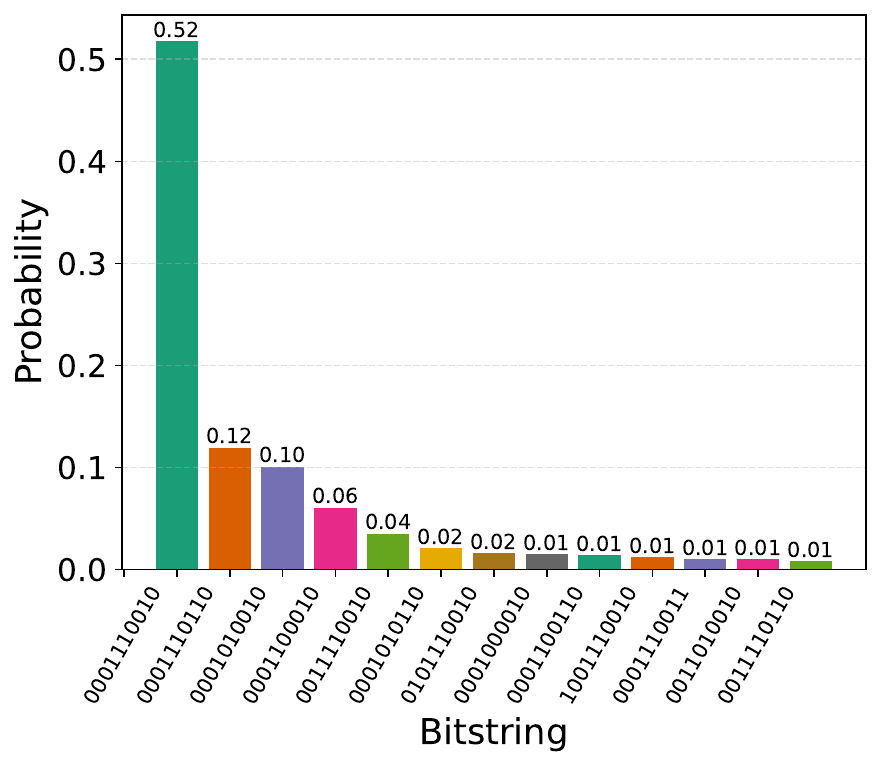}
        \par\vspace{0.5ex}
        (b)
    \end{minipage}
    \caption{Optimization results for Combination~I. (a) Evolution of the loss function value for Combination I under the optimal learning-rate configuration. (b) Final probability distribution of the quantum state for Combination I.}
    \label{fig:appendix_comb1_A2A3}
\end{figure}

Using the optimal learning-rate configuration, a representative optimization trajectory is shown in Fig.~\ref{fig:appendix_comb1_A2A3}(a). The loss function decreases monotonically with the iteration number. The corresponding probability distribution of the final quantum state is reported in Fig.~\ref{fig:appendix_comb1_A2A3}(b), where the target state achieves the highest probability.

Through computation, we identify six candidate keys corresponding to the known plaintext--ciphertext pair, namely \texttt{0000111010}, \texttt{0001110010}, \texttt{0010011111}, \texttt{0011010111}, \texttt{1011011111}, and \texttt{1100110000}. Among these, the occurrence probability of \texttt{0001110010} is $0.52$, while those of the remaining candidates are negligible. Therefore, the overall success probability, defined as the sum of these probabilities, is approximately $0.52$.

\subsection{Combination III}
Combination III corresponds to the configuration with Grover-enhanced state, a unitary Ansatz, a 7-regular graph-structured cost function, and gradient descent optimization, as summarized in Table~II.

\begin{figure}[h]
    \centering
    \includegraphics[width=0.55\linewidth]{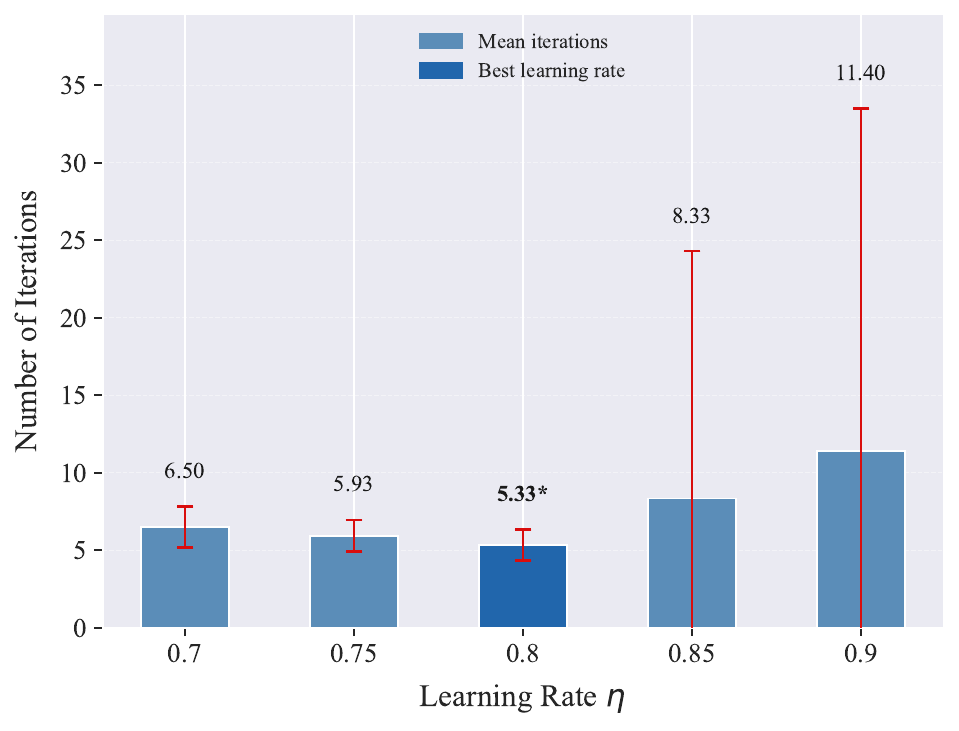}
    \caption{Average number of optimization iterations as a function of the learning rate for Combination III. Lower values indicate faster convergence.}
    \label{fig:appendix_comb3_lr}
\end{figure}

The number of Grover iterations is set to one. Following the same evaluation protocol as in Section~IV, we first determine the optimal learning rate by comparing the average number of optimization iterations under different learning-rate settings. Figure~\ref{fig:appendix_comb3_lr} shows the dependence of the iteration count on the learning rate for this configuration. The optimal learning rate is selected as the one that minimizes the required number of iterations.

\begin{figure}[H]
    \centering
    \begin{minipage}[c]{0.48\linewidth}
        \centering
        \includegraphics[width=\linewidth]{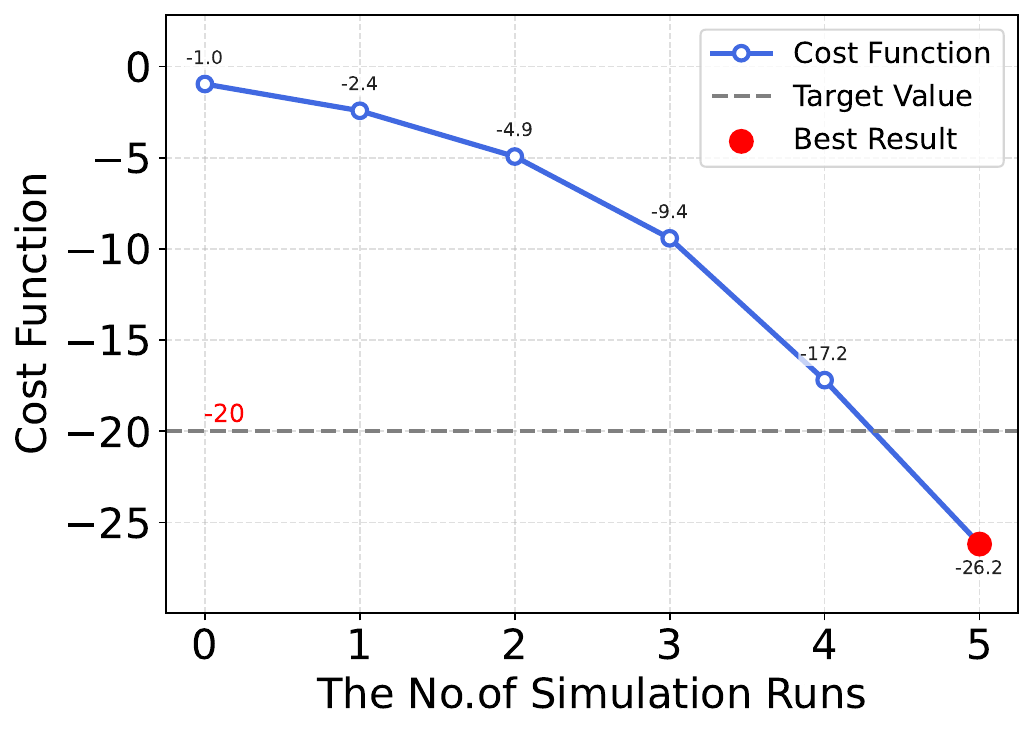}
        \par\vspace{0.5ex}
        (a)
    \end{minipage}\hfill
    \begin{minipage}[c]{0.48\linewidth}
        \centering
        \includegraphics[width=\linewidth]{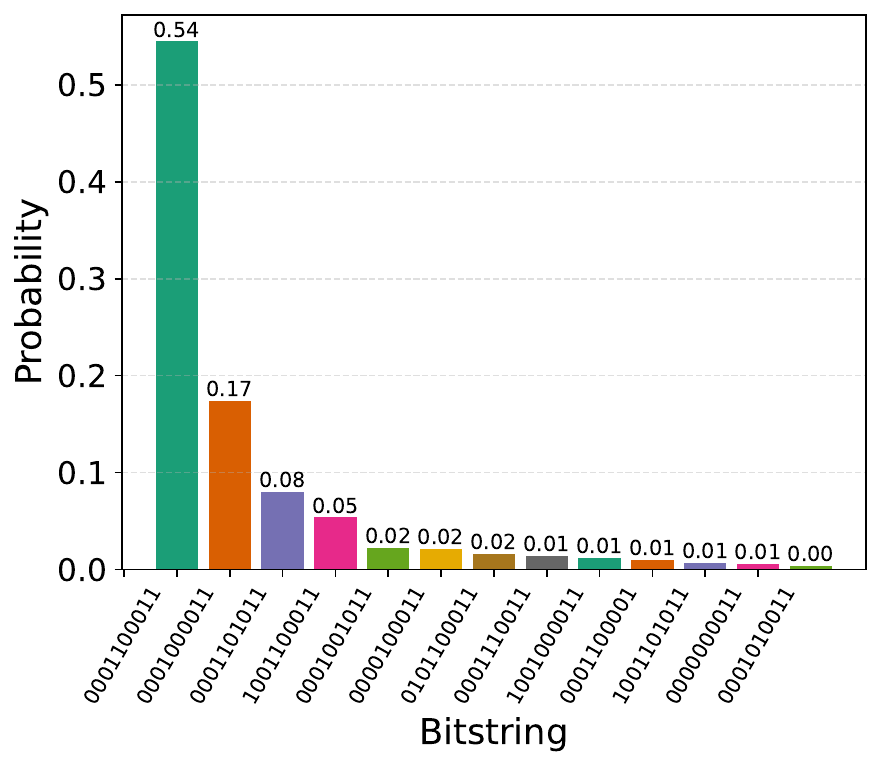}
        \par\vspace{0.5ex}
        (b)
    \end{minipage}
    \caption{Optimization results for Combination~III. (a) Evolution of the loss function value for Combination III under the optimal learning-rate configuration. (b) Final probability distribution of the quantum state for Combination III.}
    \label{fig:appendix_comb3_A2A3}
\end{figure}

Using the optimal learning-rate configuration, a representative optimization trajectory is shown in Fig.~\ref{fig:appendix_comb3_A2A3}(a). The loss function decreases monotonically with the iteration number. The corresponding probability distribution of the final quantum state is reported in Fig.~\ref{fig:appendix_comb3_A2A3}(b), where the target state achieves the highest probability.

Through computation, we identify four candidate keys corresponding to the known plaintext--ciphertext pair, namely \texttt{0001100011}, \texttt{0011100110}, \texttt{1000100011}, and \texttt{1001101011}. Among these, the occurrence probabilities of \texttt{0001100011} and \texttt{1001101011} are $0.54$ and $0.01$, respectively, while those of the remaining candidates are negligible. Therefore, the overall success probability, defined as the sum of these probabilities, is approximately $0.55$.

\subsection{Combination IV}
Combination IV corresponds to the configuration with uniform superposition initialization, a unitary Ansatz, a 0-regular graph-structured cost function, and gradient descent optimization, as summarized in Table~II.

\begin{figure}[h]
    \centering
    \includegraphics[width=0.55\linewidth]{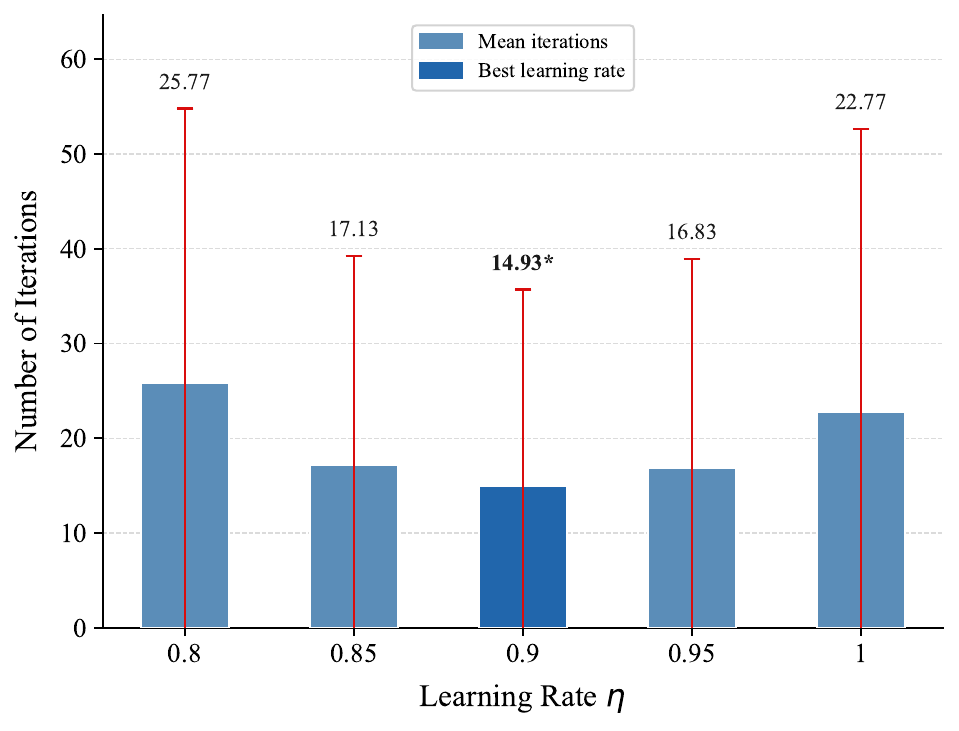}
    \caption{Average number of optimization iterations as a function of the learning rate for Combination IV. Lower values indicate faster convergence.}
    \label{fig:appendix_comb4_lr}
\end{figure}

Following the same evaluation protocol as in Section~IV, we first determine the optimal learning rate by comparing the average number of optimization iterations under different learning-rate settings. Figure~\ref{fig:appendix_comb4_lr} shows the dependence of the iteration count on the learning rate for this configuration. The optimal learning rate is selected as the one that minimizes the required number of iterations.

\begin{figure}[H]
    \centering
    \begin{minipage}[c]{0.48\linewidth}
        \centering
        \includegraphics[width=\linewidth]{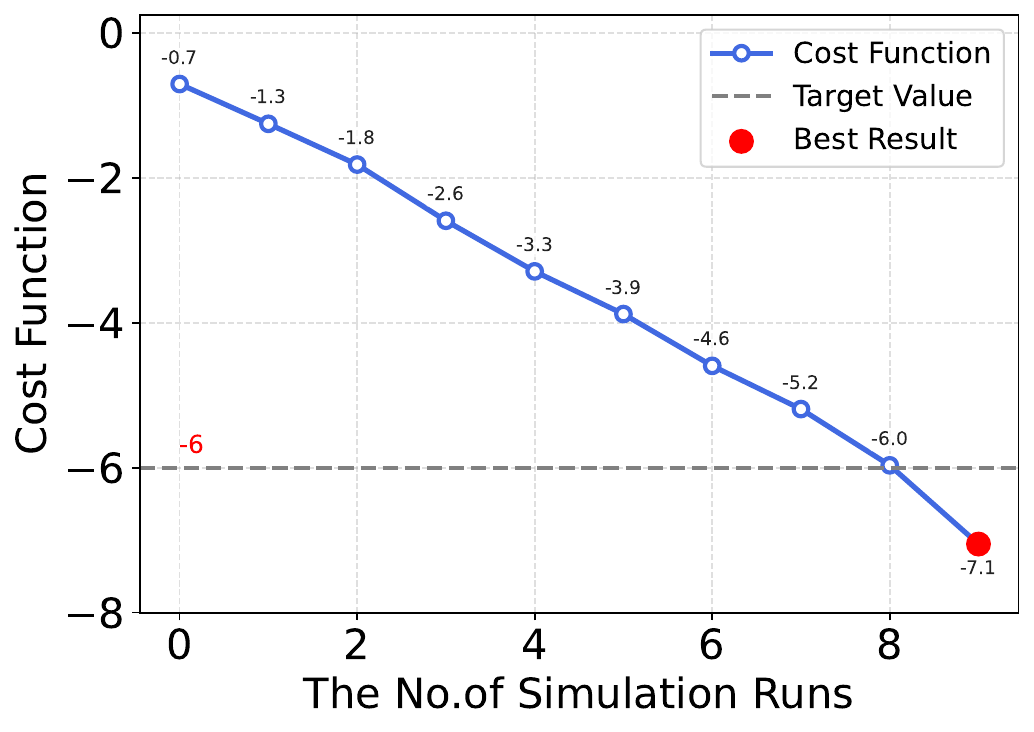}
        \par\vspace{0.5ex}
        (a)
    \end{minipage}\hfill
    \begin{minipage}[c]{0.48\linewidth}
        \centering
        \includegraphics[width=\linewidth]{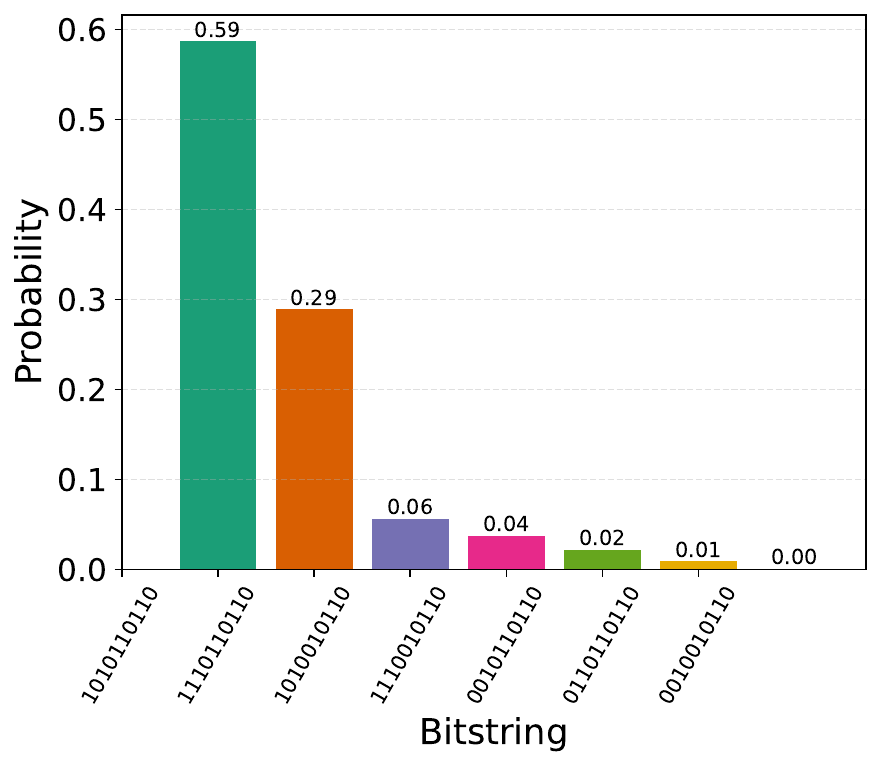}
        \par\vspace{0.5ex}
        (b)
    \end{minipage}
    \caption{Optimization results for Combination~IV. (a) Evolution of the loss function value for Combination IV under the optimal learning-rate configuration. (b) Final probability distribution of the quantum state for Combination IV.}
    \label{fig:appendix_comb4_A2A3}
\end{figure}

Using the optimal learning-rate configuration, a representative optimization trajectory is shown in Fig.~\ref{fig:appendix_comb4_A2A3}(a). The loss function decreases monotonically with the iteration number. The corresponding probability distribution of the final quantum state is reported in Fig.~\ref{fig:appendix_comb4_A2A3}(b), where the target state achieves the highest probability.

Through computation, we identify nine candidate keys corresponding to the known plaintext--ciphertext pair, namely \texttt{0010101000}, \texttt{0010111100}, \texttt{0101010011}, \texttt{0110101000}, \texttt{0110111100}, \texttt{1010100010}, \texttt{1010110110}, \texttt{1011101010}, and \texttt{1011111110}. Among these, the occurrence probability of \texttt{1010110110} is $0.59$, while those of the remaining candidates are negligible. Therefore, the overall success probability, defined as the sum of these probabilities, is approximately $0.59$.

\subsection{Combination V}
Combination V corresponds to the configuration with uniform superposition initialization, a unitary Ansatz, a 0-regular graph-structured cost function, and Nelder–Mead optimization, as summarized in Table~II.

In the Nelder--Mead (N--M) algorithm, the initial simplex is constructed as follows. 
For each parameter direction, if the initial value is zero, it is set to $1.57$; 
otherwise, the parameter is multiplied by $1.57$. 
Since all initial parameters are set to zero in our experiments, this procedure is equivalent to constructing a hypercube centered at the origin with edge length $1.57$. 
The choice of $1.57$ has a clear physical motivation: it is approximately equal to $\pi/2$, i.e., half of the period of a single-parameter rotation. 
This choice allows the subsequent N--M optimization to efficiently explore the entire periodic domain with moderate step sizes, while avoiding an excessively large initial search volume that would increase the iteration complexity.

During the optimization, we employ the following restart criterion:
\begin{equation}
\label{eq:restart}
\text{Restart if }
\begin{cases}
\left| y_{-1} - y_{0} \right| < 10^{-2}, \\[4pt]
\left| \mathrm{funvalue}_{-\mathrm{bp}} - \mathrm{funvalue}_{-1} \right| < 10^{-4},
\end{cases}
\end{equation}
where $\text{b\_p}=50$.
The condition $\left| y_{-1} - y_{0} \right| < 0.01$ indicates that the maximum variation of the loss function over all vertices of the simplex has fallen below a threshold, suggesting that the search region has contracted to a very small volume without reaching the target value, and the algorithm is likely trapped in a local minimum.
The second condition monitors the improvement of the best loss value over the previous $50$ iterations; if the change becomes negligible, the optimization is considered to have stagnated or entered a cyclic behavior, and a restart is triggered.

In addition, similar to stopping criteria commonly adopted in gradient-based methods, the optimization is terminated once the total number of measurements reaches $2^{10}=1024$.
This threshold is comparable to the complexity of a classical brute-force search and indicates that the target state is expected to have been encountered.

\begin{figure}[h]
    \centering
    \includegraphics[width=0.8\linewidth]{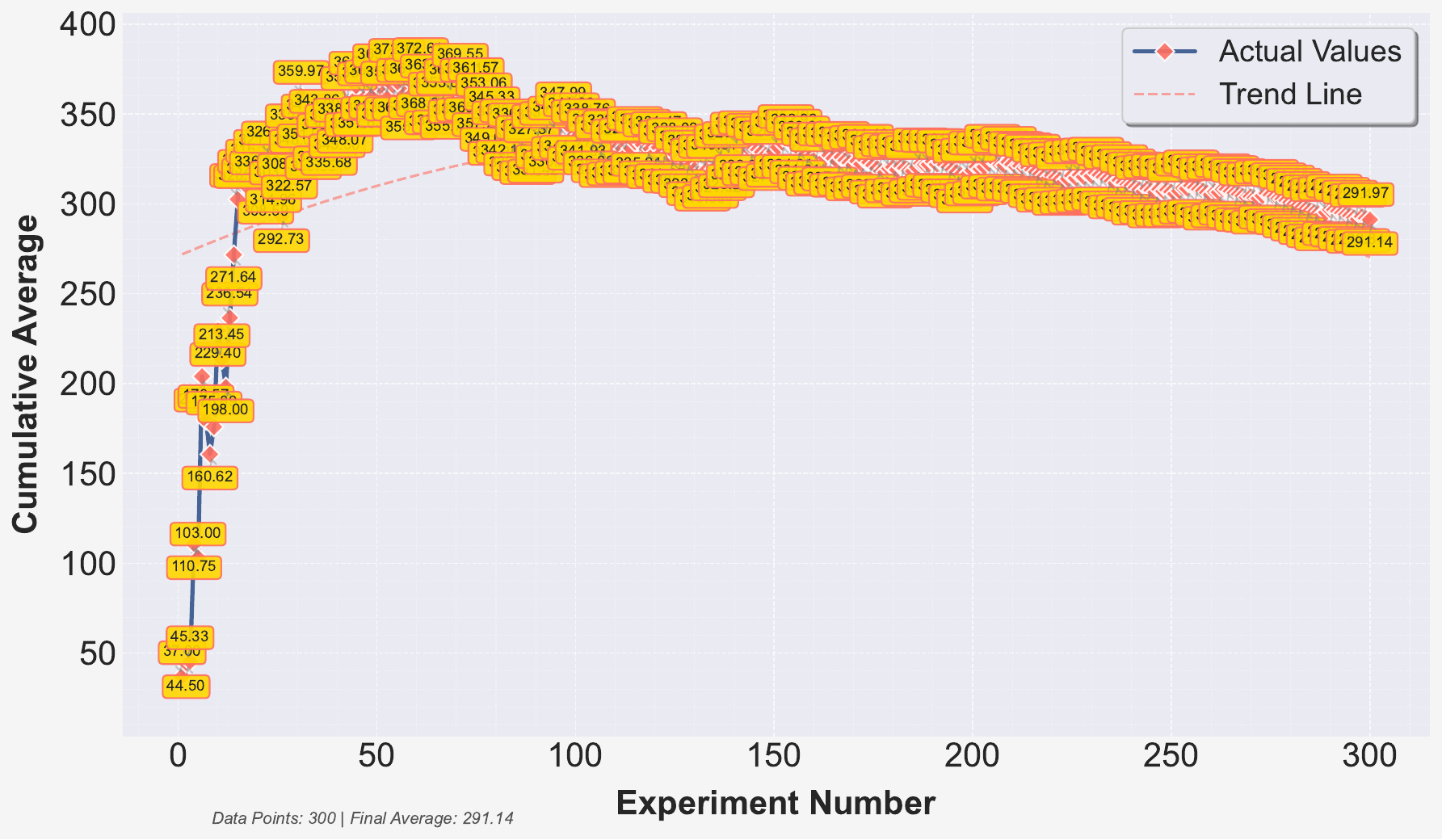}
    \caption{Average number of evaluations of the VQAA for S-DES under the N–M optimization algorithm as a function of the number of trials.}
    \label{fig:appendix_comb5_lr}
\end{figure}

Under these settings, we perform $300$ independent numerical experiments and analyze the evolution of the average number of iterations as a function of the number of trials. 
The results are summarized in Fig.~\ref{fig:appendix_comb5_lr}. As shown in Fig.~\ref{fig:appendix_comb5_lr}, the average number of iterations stabilizes at approximately 291.14.

\begin{figure}[h]
    \centering
    \begin{minipage}[c]{0.48\linewidth}
        \centering
        \includegraphics[width=\linewidth]{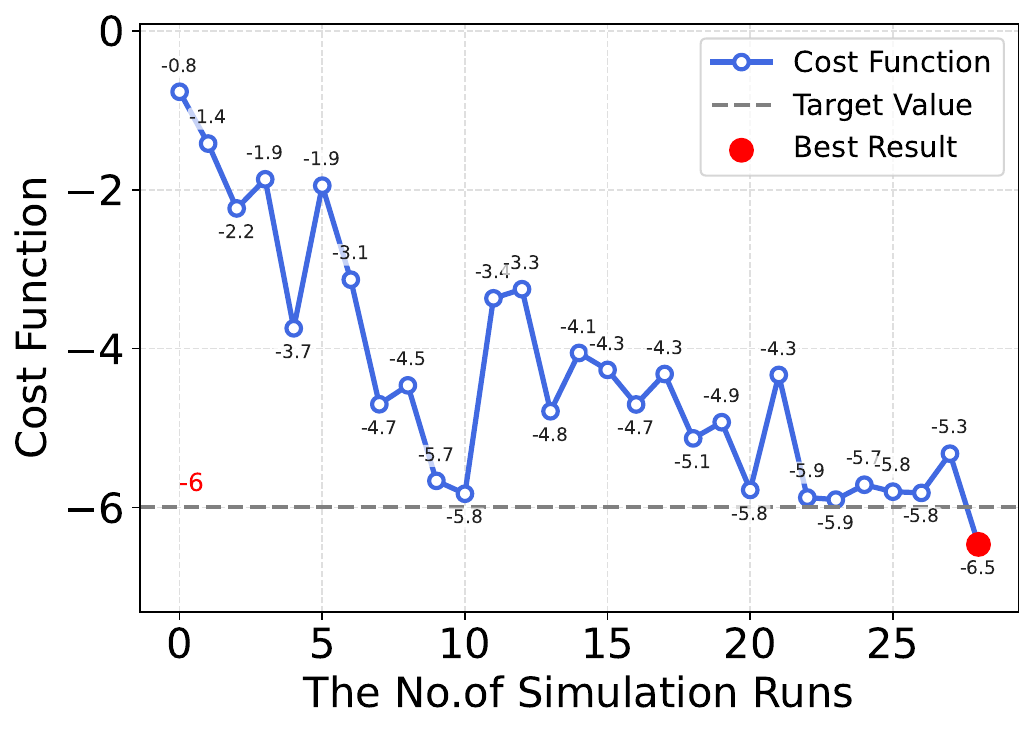}
        \par\vspace{0.5ex}
        (a)
    \end{minipage}\hfill
    \begin{minipage}[c]{0.48\linewidth}
        \centering
        \includegraphics[width=\linewidth]{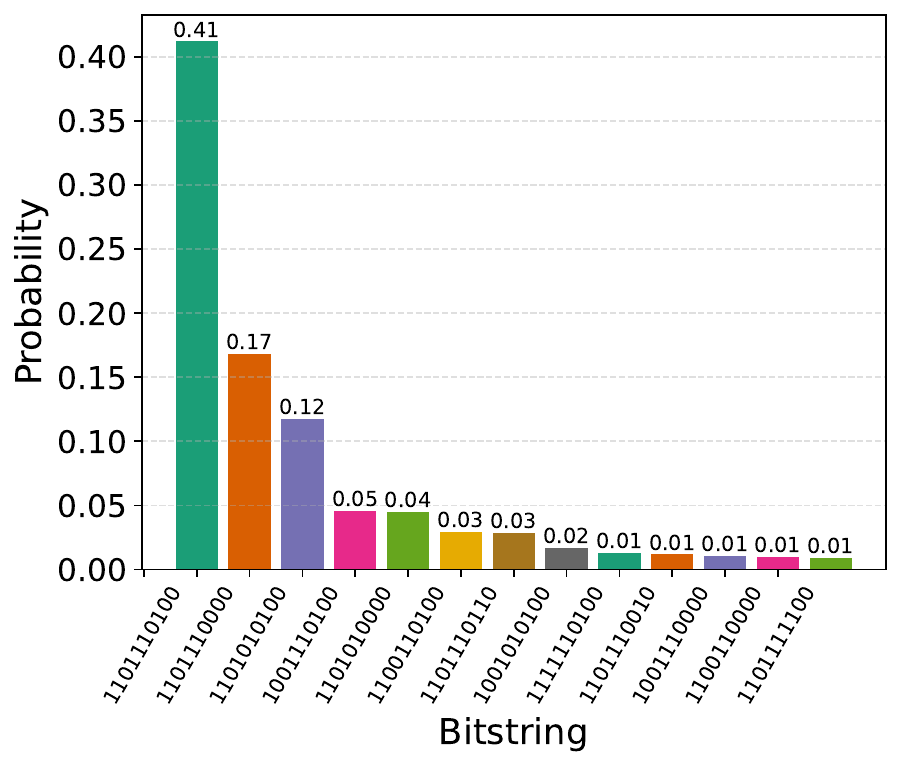}
        \par\vspace{0.5ex}
        (b)
    \end{minipage}
    \caption{Optimization results for Combination~V. (a) Evolution of the loss function value for Combination V. (b) Final probability distribution of the quantum state for Combination V.}
    \label{fig:appendix_comb5_A2A3}
\end{figure}

Using the optimal hyperparameter configuration, a representative optimization trajectory is shown in Fig.~\ref{fig:appendix_comb5_A2A3}(a). The loss function decreases monotonically with the iteration number. The corresponding probability distribution of the final quantum state is reported in Fig.~\ref{fig:appendix_comb5_A2A3}(b), where the target state achieves the highest probability.

Through computation, we identify ten candidate keys corresponding to the known plaintext--ciphertext pair, namely \texttt{0101010010}, \texttt{0101110110}, \texttt{1000011000}, \texttt{1000111100}, \texttt{1001010000}, \texttt{1001110100}, \texttt{1100011000}, \texttt{1100111100}, \texttt{1101010000}, and \texttt{1101110100}. Among these, the occurrence probabilities of \texttt{1001010000}, \texttt{1001110100}, \texttt{1101010000}, and \texttt{1101110100} are $0.01$, $0.05$, $0.04$, and $0.41$, respectively, while those of the remaining candidates are negligible. Therefore, the overall success probability, defined as the sum of these probabilities, is approximately $0.51$.

\subsection{Combination VI}
Combination VI corresponds to the configuration with Grover-enhanced initialization, a unitary Ansatz, a 0-regular graph-structured cost function, and Nelder–Mead optimization, as summarized in Table~II. The hyperparameter settings are the same as those in the previous section.

\begin{figure}[H]
    \centering
    \includegraphics[width=0.8\linewidth]{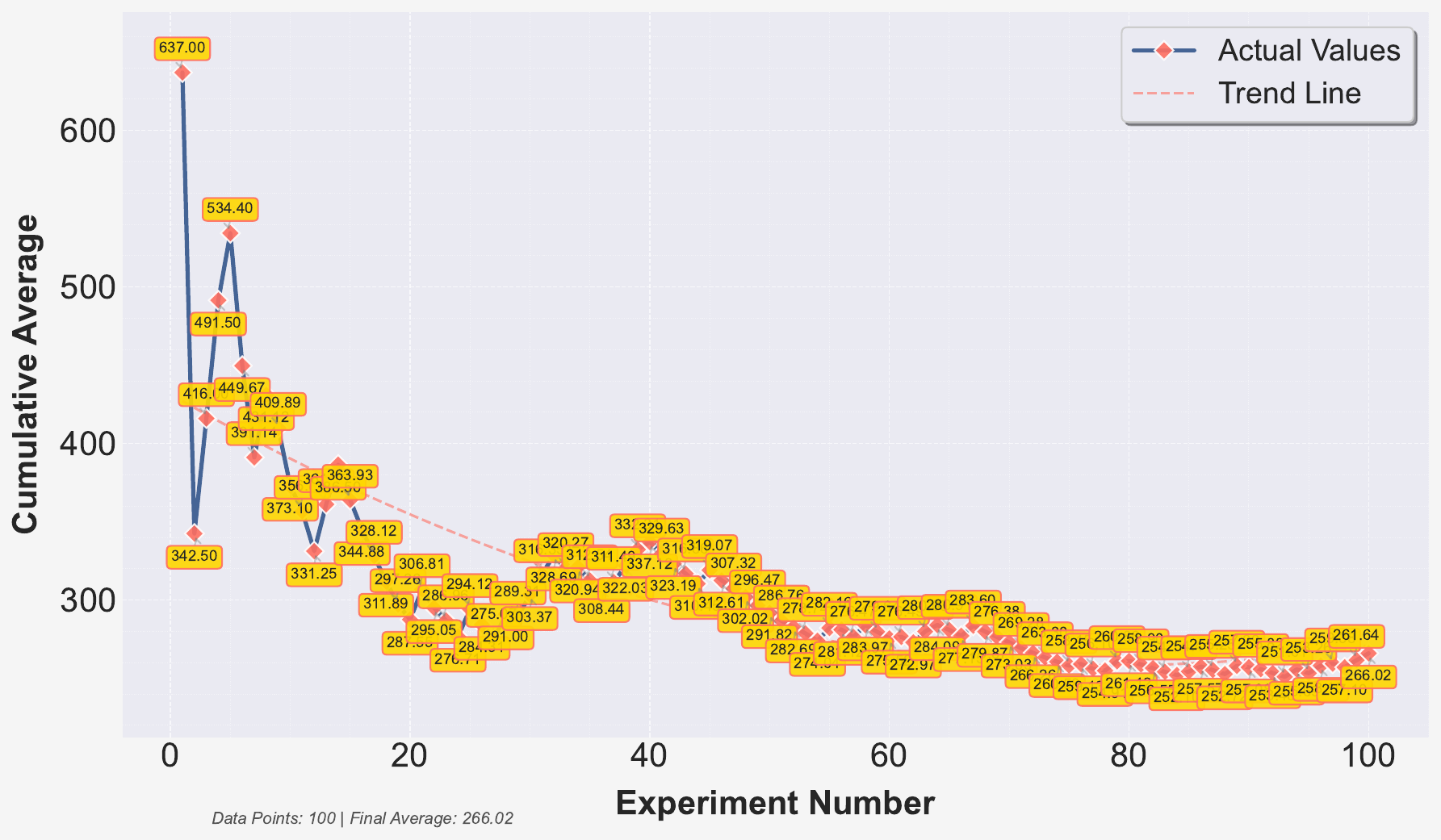}
    \caption{Average number of evaluations of the VQAA for S-DES under the N–M optimization algorithm as a function of the number of trials.}
    \label{fig:appendix_comb6_lr}
\end{figure}

Under these settings, we perform $100$ independent numerical experiments and analyze the evolution of the average number of iterations as a function of the number of trials. 
The results are summarized in Fig.~\ref{fig:appendix_comb6_lr}. As shown in Fig.~\ref{fig:appendix_comb6_lr}, the average number of iterations stabilizes at approximately 266.02.

\begin{figure}[h]
    \centering
    \begin{minipage}[c]{0.48\linewidth}
        \centering
        \includegraphics[width=\linewidth]{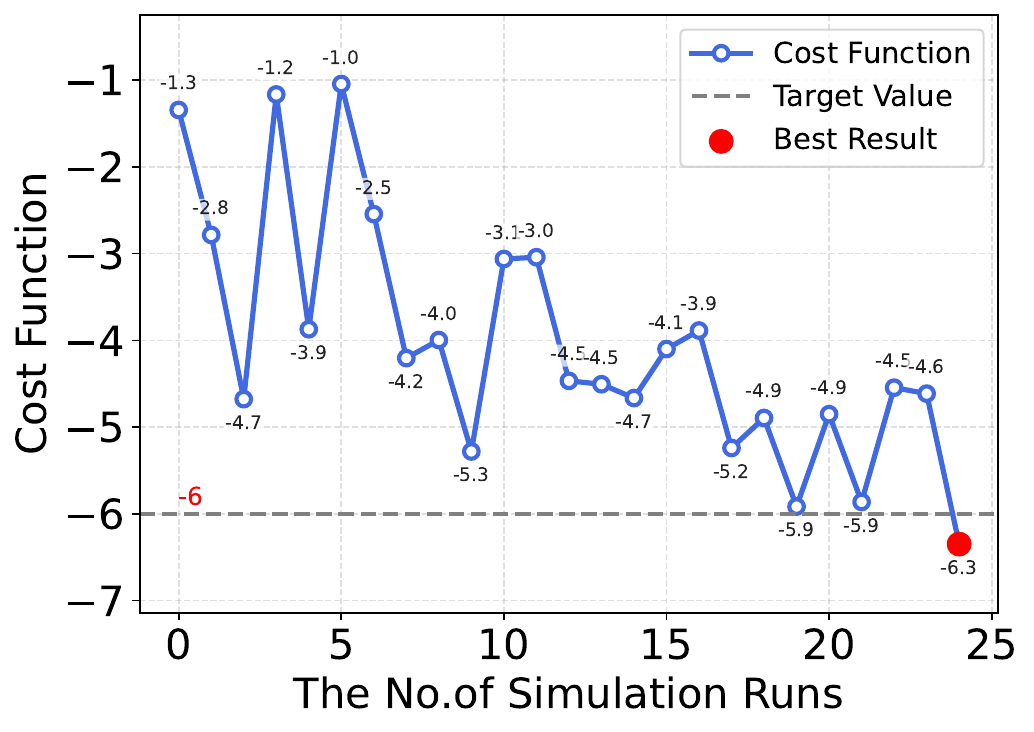}
        \par\vspace{0.5ex}
        (a)
    \end{minipage}\hfill
    \begin{minipage}[c]{0.48\linewidth}
        \centering
        \includegraphics[width=\linewidth]{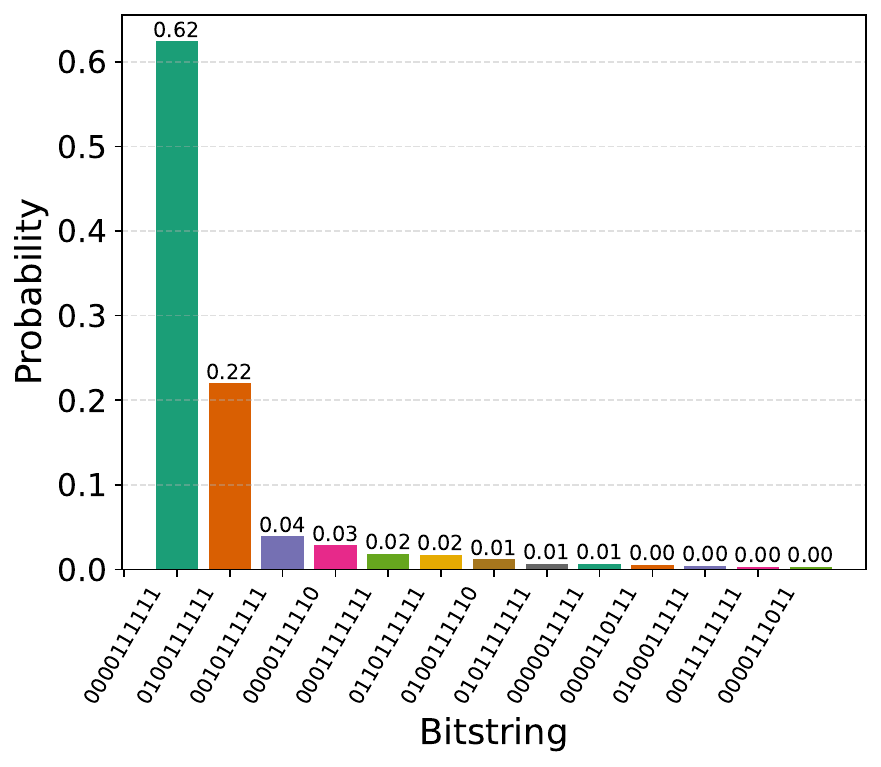}
        \par\vspace{0.5ex}
        (b)
    \end{minipage}
    \caption{Optimization results for Combination~VI. (a) Evolution of the loss function value for Combination VI. (b) Final probability distribution of the quantum state for Combination VI.}
    \label{fig:appendix_comb6_A2A3}
\end{figure}

Using the optimal hyperparameter configuration, a representative optimization trajectory is shown in Fig.~\ref{fig:appendix_comb6_A2A3}(a). The loss function decreases monotonically with the iteration number. The corresponding probability distribution of the final quantum state is reported in Fig.~\ref{fig:appendix_comb6_A2A3}(b), where the target state achieves the highest probability.

Through computation, we identify six candidate keys corresponding to the known plaintext--ciphertext pair, namely \texttt{0000111111}, \texttt{0001110111}, \texttt{0010111111}, \texttt{0011110111}, \texttt{1000111101}, and \texttt{1010111101}. Among these, the occurrence probabilities of \texttt{0000111111} and \texttt{0010111111} are $0.62$ and $0.04$, respectively, while those of the remaining candidates are negligible. Therefore, the overall success probability, defined as the sum of these probabilities, is approximately $0.66$.

\end{document}